\documentclass[10pt,reqno]{article}
\usepackage{amsmath,amssymb,amsthm}
\usepackage{esint}
\usepackage{bm}
\usepackage{url}
\usepackage{subfigure}
\usepackage{graphicx,color}
\usepackage{algorithm}
\usepackage{algpseudocode}
\usepackage{algorithmicx}
\usepackage{hyperref}
\usepackage{float}
\usepackage{booktabs,multirow}
\usepackage{hhline}
\usepackage{setspace}
\usepackage{subfigure}
\usepackage{xcolor}
\usepackage{lipsum}
\usepackage{wasysym}  
\usepackage{caption}
\usepackage{hyphsubst}
\usepackage{filecontents}
\usepackage{hyperref}

\hypersetup{colorlinks,citecolor=blue,linkcolor=blue, urlcolor=blue}
\usepackage{pdflscape}
\usepackage{breakurl}
\newcommand{\RR}{\mathbb{R}}
\newcommand{\J}{\jmath}

\newcommand{\ds}{\displaystyle}

\DeclareMathAlphabet{\itbf}{OML}{cmm}{b}{it}

\newcommand{\email}[1]{\protect\href{mailto:#1}{#1}}

\newcommand{\pathfigures}{Figures/}
\graphicspath{{\pathfigures}}

\date{\today} 
\begin{document}
	
	\title{Acoustic scattering from a wave-bearing cavity with flexible inlet and outlet
		%\thanks{ABC EFG}
	}
	
	\author{
		Muhammad Afzal\footnotemark[1]
		\and
		Hazrat Bilal\footnotemark[1] 
		\and
		Naveed Ahmed\footnotemark[2]
		\and 
		Abdul Wahab\footnotemark[3] \footnotemark[4] 
	}
	\maketitle
	
	\renewcommand{\thefootnote}{\fnsymbol{footnote}}
	\footnotetext[1]{Department of Mathematics, Capital University of Science and Technology, Islamabad, 46000, Pakistan (dr.mafzal@cust.edu.pk;  bilalqau77@gmail.com)}
	\footnotetext[2]{Gulf University for Science \& Technology, Block 5, Building 1, Mubarak Al-Abdullah Area, West Mishref, Kuwait
(ahmed.n@gust.edu.kw)}
	\footnotetext[3]{Department of Mathematics, School of Sciences and Humanities, Nazarbayev University, 53, Kabanbay Batyr Avenue, 010000, Astana, Kazakhstan (abdul.wahab@nu.edu.kz)}
	\footnotetext[4]{Address all correspondence to A. Wahab at   \email{abdul.wahab@nu.edu.kz}.}
	\renewcommand{\thefootnote}{\arabic{footnote}}
\begin{abstract} 
In this article, we substantiate the appositeness of the \emph{mode-matching technique} to study the scattering response of bridging elastic plates connecting two flexible duct regions of different heights. We present two different solution schemes to analyze the structure-borne and fluid-borne radiations in the elastic plate-bounded waveguide. The first scheme supplements the mode-matching technique with the so-called \emph{tailored-Galerkin approach} which uses a solution ansatz with homogeneous and integral parts corresponding to the vibrations of the bridging elastic plate and the cavity, respectively. In the second scheme, we supplement the mode-matching technique with the \emph{modal approach} wherein the displacement of the bridging elastic plate is projected onto the eigenmodes of the cavity. To handle the non-orthogonality of the eigenfunctions, we invoke generalized orthogonality relations. An advantage of the proposed mode-matching schemes is that they provide a convenient way of incorporating a variety of edge conditions on the joints of the plates, including clamped, pin-joint, or restraint connections. The numerical analysis of the waveguide scattering problems substantiates that edge connections on the joints have a significant impact on the scattering energies and transmission loss. 
\end{abstract}


\section{Introduction}

The problems involving wave scattering at structural discontinuities are challenging as well as intriguing. Because of their numerous applications, they have piqued the interest of engineers and scientists. Abrupt geometric variations, changes in the material properties of the medium and bounding walls, and the varying conditions on joints or edges in acoustic and electromagnetic waveguides significantly influence the propagation and scattering of waves \cite{Doak, Dowell, Pan, Huang, ammari1, ammari2}. Such characteristics are pertinent, for instance, for designing active and passive noise control devices such as silencers, where the aim is to reduce pressure fluctuations or attenuate unwanted noise. These structural discontinuities also play a critical role in the designs of solid structures to reduce the vibrations generated by engines, air flows, and gearboxes and the low-frequency vibrations of thin elastic plates that cover the structural components in submarines \cite{Ming, Lau1, Du, Lawrie12, Kim, Mimani}.

To understand how energy propagates in complex structures and scatters at abrupt variations, flanges, and joints, it is necessary to model and analyze fundamental problems. The current research focuses on a fluid-structure coupled wave scattering problem with elastic plates bounding a fluid matrix. The geometric structure has multiple joints and scatters. Note that the reflection and transmission of waves at a planar interface of two plates having different material properties are well understood; see, for instance, \cite{Norris}. However, the situation becomes complicated if the plates are immersed in a fluid, given the leakage of structural energy at the edges of the fluidic medium. An important acoustic scattering problem involving the diffraction of the acoustic waves from a semi-infinite plate embedded in a fluid was considered by Cannell \cite{Cannell}. He used the Wiener-Hopf technique to solve the resulting boundary value problem. Similar studies were conducted, for instance, in \cite{Lawrie, Grant, Thompson, Warren}.

A framework for deriving the analytic solutions of boundary value problems consisting of the Helmholtz equation and high-order boundary conditions (for elastic membranes and elastic plates) subject to various edge conditions is offered by Lawrie and Abraham \cite{Law1999}. Their technique, coined the \textit{mode-matching technique} (MMT), relies on a discrete description of the spectrum rather than a continuous formulation, in contrast to the integral transform-based approaches. It provides a convenient way to work with a solution ansatz in terms of the eigenfunctions of the associated eigenvalue problems. It is important to mention that the eigenfunctions corresponding to the elastic membranes and elastic plates (with high-order boundary conditions) are non-orthogonal and linearly dependent. Therefore, the so-called generalized orthogonality relations are indispensably employed to achieve a convergent solution, see, for instance, \cite{Lawrie2007}. For a variety of interesting problems catered for using the MMT, we refer interested readers to \cite{Hassan09, Afzal16, Afzal16b, Nawaz21, Afzal21, Bilal22, Aqsa, Sumbul}. Recently, Nawaz and Lawrie \cite{RN13Jasa} used the MMT for a scattering problem involving a rigid-soft flange at the discontinuous interface of two semi-infinite elastic plates bounding a fluid matrix. The approach was followed in \cite{Afzal20Jasa, AfzalCNSNS21, Afzal22Jasa} to analyze the impact of acoustic linings along the flanges of a flexible expansion chamber. 

The current study builds upon the ideas presented by Lawrie and Afzal \cite{LawrieAfzal} to analyze the acoustic scattering from a bridging membrane-height discontinuity at the interface of an infinite membrane-bounded waveguide. The MMT was supplemented with Galerkin and tailored-Galerkin approaches to derive the solution to the scattering problem. The Galerkin method is based on the assumption of prior solutions that change depending on the conditions at the bridging membrane's edges. In the tailored-Galerkin approach, the fluid-structure coupled modes of a duct region are used to model the vibration response along the bridging membrane. A unique representation of membrane displacement in the advanced setting can reveal the scattering response with a variety of edge conditions. The aforementioned problem is a prototype set involving two semi-infinite sections bounded by elastic membranes and bridging membrane heights at the interface. Thus, the practical configuration comprises a finite-length expansion chamber that includes two bridging-height discontinuities, elastic plates, and sets of edge conditions.

The model problem is physically interesting given the consideration of elastic plates together with a variety of edge conditions and their effects on scattering. The elastic plates can be connected in a variety of ways. For instance, if the plates are connected with a clamped connection, the displacements and their gradients are zero at the joint. On the other hand, if the plates are pin-jointed, the displacement and bending moment are zero at the joint. Thus, each physical setup yields a different set of mathematical conditions. In this article, the Galerkin and tailored-Galerkin approaches are used to study their responses with more advanced formulations. 

The article is organized as follows. The waveguide structure, governing equations, and idea of the MMT are presented in Section \ref{sec:Form}. The MMT solution is supplemented with the tailored-Galerkin approach in Section \ref{sec:TG} and with the modal approach in Section \ref{sec:Modal}. The numerical results are presented and discussed in Section \ref{sec:Num}. The findings of the article are summarized in Section \ref{sec:Con}.
	
\section{Mathematical formulation}\label{sec:Form}  

In this section, we present the geometric structure of the waveguide and mathematically formulate the related scattering problem. We also introduce some notation and give some preliminary material beforehand to facilitate the ensuing study.

\subsection{Waveguide structure}

\begin{figure}[htb!]
\begin{center}
\includegraphics[width=.6\textwidth,  height=0.4\textwidth]{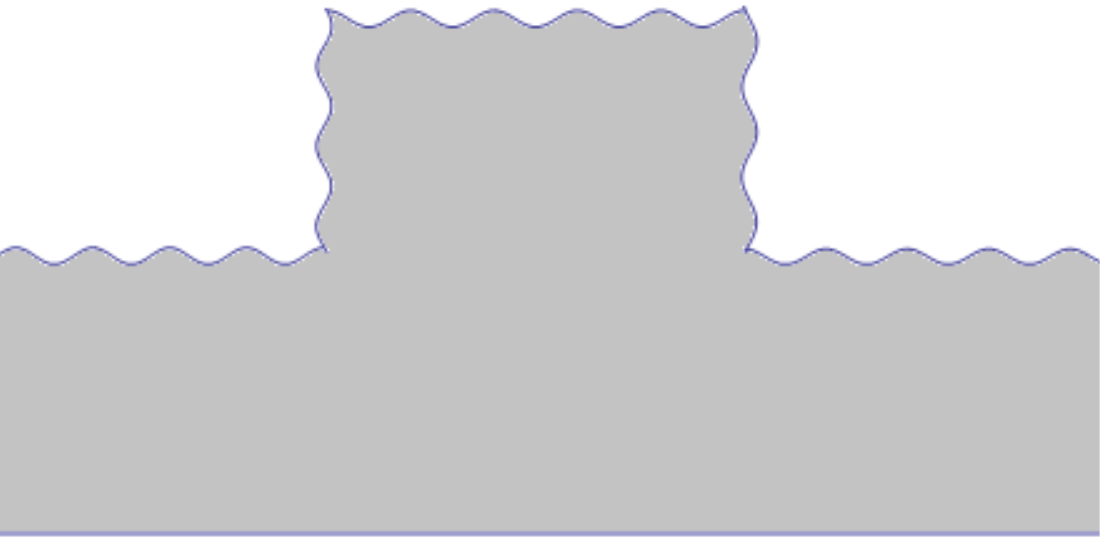}
\caption{Structural configuration  of  the  waveguide. The wavy  boundaries  represent  elastic plates.}\label{fig:1}
\end{center}
\end{figure}

In this article, we consider an infinite two-dimensional rectangular waveguide with an expansion chamber stretched along $\bar{x}-$axis in the $(\bar{x},\bar{y})$ plane. The lower wall of the waveguide ($\overline{\mathbb{W}}_0:=\mathbb{R}\times\{\bar{0}\}$) is acoustically rigid, but the upper walls ($\overline{\mathbb{W}}_1:=(-\infty,-\bar{L})\times\{\bar{a}\}$, $\overline{\mathbb{W}}_2:=(\bar{L},\infty)\times\{\bar{a}\}$, and $\overline{\mathbb{W}}_3:=(-\bar{L},\bar{L})\times\{\bar{b}\}$, for some fixed $\overline{L}\in\RR^+$) consist of elastic plates. The waveguide is divided into three duct regions (the inlet, the expansion chamber, and the outlet) by two vertical elastic strips along the line segments $\{\bar{L}\}\times[\bar{a},\bar{b}]$ and $\{ -\bar{L}\}\times[\bar{a},\bar{b}]$. The waveguide is loaded with a compressible fluid having density $\rho$ and speed of sound $c$ whereas the outside of the waveguide is assumed to be \textit{in-vacuo}. The structural configuration of the waveguide is depicted in Fig. \ref{fig:1}. In the above and what follows, a superposed bar indicates a dimensional quantity, and we will drop the bar for non-dimensional counterparts.

\subsection{Governing equations} 

The waveguide structure is excited by a duct mode from the region $(-\infty,-\overline{L})\times(0,\bar{a})$ that propagates from negative to positive $\bar{x}$-direction. The acoustic pressure, acoustic velocity, and the fluid potential inside the waveguide ($\bar{p}$, $\bar{\mathbf{v}}$, and $\bar{\Phi}$, respectively) satisfy the partial differential equations
\begin{align}
\bar{p} =  -\rho  \dfrac{\partial\bar{\Phi}}{\partial\bar{t}}
\quad\text{and}\quad
\bar{\mathbf{v}}=\bar{\nabla}  \bar{\Phi}.
\label{eq:1}
\end{align}  
Assuming a time-harmonic dependence with angular frequency $\omega$ and wavenumber $k:=\omega/c$, we define length scale $k^{-1}$ and time scale $\omega^{-1}$ so that  $x:=k\bar{x}$,  $y:=k\bar{y}$,  and  $t:=\omega\bar{t}$ become non-dimensional space and time coordinates. Thanks to the relations in Eq. \eqref{eq:1}, the  time-harmonic non-dimensional  fluid  potential,  $\phi(x,y,\omega)$,  satisfies  the  Helmholtz  equation,
\begin{equation}
\nabla^2\phi  +\phi=0.
\label{eq:a3}  
\end{equation}
In the rest of this article, we do not mention the dependence on $\omega$ for ease of notation. We complement Eq. \eqref{eq:a3} with the non-dimensional boundary conditions,  
\begin{align}
\begin{cases} 
\dfrac{\partial\phi}{\partial y}=0, &\mathbb{W}_0,
\\
\dfrac{\partial^5\phi}{\partial  x^4\partial y}-\mu^4\dfrac{\partial  \phi}{\partial  y}-\alpha\phi  =  0,  &\mathbb{W}_1\cup \mathbb{W}_2\cup \mathbb{W}_3,
\end{cases}
\label{eq:5}
\end{align}
for the acoustically rigid lower wall and elastic plate upper walls, respectively.  The  quantities  
\begin{align}
\mu:=\left(\dfrac{12(1-\nu^2)c^2  \rho_p}{k^2h^2  E  }\right)^{1/4}  
\quad\text{and}\quad  
\alpha =:  \dfrac{12(1-\nu^2)c^2  \rho}{k^3h^3E },
\label{eq:51}
\end{align}
are the \textit{in-vacuo} plate wavenumber and fluid loading parameter,  respectively.  Here,  $E$, $\rho_{p}$, $\nu$, and $h$, respectively, are Young's modulus, plate density, Poisson's ratio,  and the thickness of the plates.  

\subsection{Elements of spectral theory}

It is easy to verify from Eqs. \eqref{eq:a3} and \eqref{eq:5}, that the eigenvalues to the problems in the regions $\{(x,y)\in\RR^2|\,\, L<|x|, \, 0\leq y\leq a\}$ (inlet and outlet) and $\{(x,y)\in\RR^2|\,\, |x|<L, \, 0\leq y\leq b\}$ (expansion chamber) are the roots of the dispersion relation, 
\begin{align}
\left((\varrho^{2}+1)^2-\mu^{4}\right)  \varrho  \sinh  (\varrho  \zeta)-\alpha\cosh(\varrho  \zeta)  =  0,  \label{eq:7}
\end{align}
where $\varrho=\tau$ (resp. $\varrho=\gamma$) when  elastic  plates  lie  along  $\zeta=a$  (resp. $\zeta=b$). These roots are numerically obtained and the computation of all the roots is essential for a successful implementation of the MMT.  We refer the reader to  \cite{Lawrie2007} for more details on the spectrum of the subproblems. 

The  corresponding eigenfunctions, $Y_\J(\varrho_{n},y):= \cosh(\varrho_{n}y)$,  in the inlet/outlet (with $\J=1$, $\varrho_n=\tau_n$ and $\zeta=a$) and the expansion chamber (with $\J=2$, $\varrho_n=\gamma_n$, and $\zeta=b$) are linearly dependent and non-orthogonal \cite{Lawrie2007}. Indeed, 
\begin{align}
&\ds\sum_{n=0}^{\infty} \Delta_{\J,n}Y_{\J}(\varrho_n,y)=0, \qquad \ds\sum_{n=0}^{\infty}   \varrho^2_n\Delta_{\J,n} Y_{\J}(\varrho_n,y)=0,  \quad 0\leq  y  \leq  \zeta,
\label{eq:g1}
\\
&\ds\sum_{n=0}^{\infty}  \Delta^{2}_{\J,n}\Gamma_{\J,n}=0, 
\qquad\qquad
\ds\sum_{n=0}^{\infty}  \varrho^2_{n}\Delta_{\J,n}^{2}\Gamma_{\J,n}=1. 
\label{eq:g2}
\end{align}
Here,
\begin{align}
\Gamma_{\J,n}:=\dfrac{\alpha  \zeta}{2}  +  \dfrac{\alpha  Y_\J(\varrho_n,\zeta)  Y'_\J(\varrho_n,\zeta)}{2  \varrho_n^2} +2  \left[(\varrho_n^2+1)Y'_\J(\varrho_n,\zeta)\right]^{2}
\quad\text{and}\quad 
\Delta_{\J,n}:=\dfrac{Y'_\J(\varrho_{n},\zeta)}{\Gamma_{\J,n}},
\label{eq:9}
\end{align}
where a superposed prime indicates differentiation. Accordingly, the use of generalized orthogonality relations is indispensable  to procure a convergent solution.  To that end,  the  generalized  orthogonality  relation
\begin{align}
\alpha  \int_{0}^{\zeta}Y_{\J}(\varrho_{n},y)Y_\J(\varrho_{m},y)dy=\Gamma_{\J,n}\delta_{m,n}-  (\varrho^2_{m}+\varrho^2_{n}+2)Y'_\J(\varrho_{m},\zeta)Y'_\J(\varrho_n,\zeta),
  \label{eq:8}
\end{align}
is useful, wherein $\delta_{m,n}$ is the Kronecker's delta. Moreover, we have the identity
\begin{equation}
\alpha \sum_{n=0}^{\infty}\dfrac{Y_\J(\varrho_{n},\nu)Y_\J(\varrho_{n},y)}{\Gamma_{\J,n}}=\delta_0(y+\nu)+\delta_0(y-\nu) +\delta_0(y+\nu-2\zeta), \qquad 0 \leq y, \quad \nu \leq b,
\label{Gf}
\end{equation} 
where $\delta_0$ is the Dirac mass at the origin; see, e.g., \cite{Lawrie2007}.

It is worth mentioning that the eigenfunction expansion representation of a sufficiently smooth function converges point-wise \cite{lawrie2012comments}. We also emphasize that the MMT is capable of catering to strong singularities appearing at the corners of the vertical elastic plates in velocity flux, as comprehensively discussed in \cite{RN13Jasa}.
  
\subsection{Mode matching technique}\label{sec:MMT}

We assume an ensemble time-harmonic fluid  potential ansatz of the form
\begin{eqnarray}
\phi(x,y):=
\left\{
\begin{array}{lll}
\phi_1(x,y),  &x  \leq  -L,  &0  \leq  y  \leq  a,  
\\
\phi_2(x,y),  &\vert  x  \vert  \leq  L, &0\leq  y  \leq  b,
\\
\phi_3(x,y),  &x  \geq  L, &0\leq  y  \leq  a,
\end{array}
\right.
\label{eq:10a}    
\end{eqnarray}
in terms of  the  inlet,  expansion  chamber,  and  outlet fluid potentials, $\phi_1$, $\phi_2$, and $\phi_3$, respectively. Note that these fluid  potentials can be expanded in terms of the eigenfunctions as
\begin{align}
\phi_{1}(x,y)&=F_{\ell}  Y_{1}({\tau_{\ell}},y)  e^{i\xi_{\ell}(x+L)}+  \sum_{n=0}^{\infty}  A_{n}  Y_{1}({\tau_{n}},y)e^{-i\xi_{n}(x+L)},
\label{eq:11aa}
\\
\phi_{2}(x,y)&=\sum_{n=0}^{\infty}  \left(  B_{n}e^{is_{n}x}  +  C_{n}e^{-is_{n}x}\right)Y_{2}({\gamma_{n}},y),
\label{eq:12}
\\
\phi_{3}(x,y)&=\sum_{n=0}^{\infty}  D_{n}  Y_{1}({\tau_{n}},y)e^{i\xi_{n}(x-L)},
\label{eq:13aa}
\end{align}
where $A_n$, $B_n$, $C_n$, and $D_n$ are the complex amplitudes of the reflected and transmitted modes and are to be determined.  The first term on the right-hand side of  Eq. \eqref{eq:11aa}  is the incident field with force  $F_\ell$. The index $\ell$ suggests whether the incident field consists of a fundamental mode or a higher-order mode.  The quantities    $\xi_{n}$  and  $s_{n}  $ are the wavenumbers  and are defined  in terms  of the eigenvalues $\tau_n$ and $\gamma_n$ as   
$$
\xi_{n}:= \left(\tau^{2}_{n}+1\right)^{1/2}
\quad \text{and}\quad 
s_{n}:=\left(\gamma^{2}_{n}+1\right)^{1/2}, \qquad n\in\mathbb{Z}, \,\, n\geq 0.
$$  
Further, the non-dimensional vertical elastic plate displacements, say $w^\pm(y)$, satisfy 
\begin{align}
\frac{d^4 w^\pm}{d y^4}-\mu^4 w^\pm=\pm \alpha \phi_2,\quad  \{\pm L\}\times[a,b], \label{d19}
\end{align}
in terms of the solution ansatz \eqref{eq:10a}.

The goal in the MMT is to find the amplitudes $A_n$, $B_n$, $C_n$, and $D_n$ using the continuity of the pressure flux and the normal velocities at the interfaces $x=\pm L$. To that end, the continuity of the pressure flux at the interfaces furnishes
\begin{align}
\int_0^a\phi_1(-L,y)Y_1(\tau_m,y)dy&=\int_0^a\phi_2(-L,y)Y_{1}(\tau_m,y)dy,
\label{eq:21a}  
\\
\int_0^a\phi_3(L,y)Y_1(\tau_m,y)dy&=\int_0^a\phi_2(L,y)Y_1(\tau_m,y)dy, 
\label{eq:21aa}  
\end{align}
whereas the continuity of the normal velocity flux at the interfaces yields
\begin{align}
\int_0^b\phi_{2x}(-L,y)Y_2(\gamma_m,y)  dy=&\int_0^a\phi_{1x}(-L,y)Y_2(\gamma_m,y)  dy+\int_a^bw^-(y)Y_2(\gamma_m,y)  dy,
\label{eq:14aa}  
\\
\int_0^b\phi_{2x}(L,y)Y_2(\gamma_m,y)  dy=&\int_0^a\phi_{3x}(L,y)Y_2(\gamma_m,y)  dy+\int_a^b w^+(y)Y_2(\gamma_m,y)  dy.
\label{eq:14aaa}  
\end{align}

Equations \eqref{eq:21a}-\eqref{eq:21aa} relate the acoustic pressures of the inlet and outlet regions to the acoustic pressures of the central region at $x=\pm L$.  Together with the generalized orthogonality relation \eqref{eq:8}, Eqs. \eqref{eq:21a}-\eqref{eq:21aa} render the relations 
\begin{align}
\Psi^\pm_m&=-  F_\ell\delta_{m,\ell}+\left[U_1^\pm+(\tau_m^2+2)  U_2^\pm\right] \Delta_{1,m} -\dfrac{2i\alpha}{\Gamma_{1,m}}\sum_{n=0}^{\infty}\Pi^\mp_n R_{m,n}  \chi^\pm_n,
\label{eq:19}
\end{align}
between the symmetric and anti-symmetric propagating mode amplitudes,
\begin{align}
\Psi_n^{\pm}:=A_n\pm D_n
\,\,\text{and}\,\,
\chi_n^\pm:=B_n\pm C_n,
\label{chi-psi}
\end{align}
of the inlet/outlet regions and the central region, respectively. 
Here,
\begin{align}
\Pi_n^+&:= \sin(s_n L)
\quad\text{and}\quad 
\Pi_n^-:= i\cos(s_n L), 
\\
R_{m,n}&:=\int_0^a\cosh(\tau_my)\cosh(\gamma_ny)dy,
\quad
U_1^{\pm}:=e_1\pm  e_3, 
\quad\text{and}\quad 
U_2^{\pm}:=e_2\pm e_4,\label{eq:r1}
\end{align} 
with 
\begin{eqnarray}
\left\{
\begin{array}{ll} 
e_1:= \phi_{1yyy}(-L,a), 
&e_2:= \phi_{1y}(-L,a), 
\\
e_3:=\phi_{3yyy}(L,a),
&e_4:=\phi_{3y}(L,a).
\end{array}
\right.
\label{e14}
\end{eqnarray} 
We precise that $U_j^\pm$, for $j=1,2$, are auxiliary unknowns to be determined later on. We refer the reader to Appendix \ref{append:a} for the derivation of expressions \eqref{eq:19}. The amplitudes of the modes propagating in the positive and negative directions of the waveguide can be procured subsequently as
\begin{align*} 
A_m=\dfrac{1}{2}(\Psi_m^+ +\Psi_m^-), \quad D_m=\dfrac{1}{2}(\Psi^+_m - \Psi^-_m), \quad 
 B_m=\dfrac{1}{2}(\chi_m^+ +\chi_m^-), \quad C_m=\dfrac{1}{2}(\chi_m^+ - \chi_m^-).
\end{align*}

Before using the continuity of the normal velocities to derive another set of equations for $\chi^\pm_n$ and $\Psi^\pm_n$, it is important to cater for the effects of the vertical elastic plates to uniquely solve Eq. \eqref{d19} for $w^\pm$. Towards this end, we supplement the MMT with two different approaches. We use a tailored-Galerkin approach in Section \ref{sec:TG} and a model approach in Section \ref{sec:Modal}.

\section{Tailored-Galerkin approach}\label{sec:TG}

The tailored-Galerkin approach is based on an assortment of trial functions that determine the vibrational response of the incident energy along the vertical boundaries and their edges. We look for the solutions to Eq. \eqref{d19} in the form
\begin{align}
w^-(y)=&a_1\cos(\mu y)+a_2 \sin(\mu y)+a_3\cosh(\mu y)+a_4\sinh(\mu y)
\nonumber \\ 
&-\alpha \sum_{n=0}^{\infty} \frac{\left( B_n e^{-is_n L}+C_n e^{is_n L}\right)\cosh(\gamma_n y)}{\gamma_n^4-\mu^4}, \label{d21}  
\\
w^+(y)=& a_5 \cos(\mu y)+ a_6 \sin(\mu y)+a_7\cosh(\mu y)+a_8\sinh(\mu y)
\nonumber \\
&+\alpha \sum_{n=0}^{\infty} \frac{\left( B_n e^{is_n L}+C_n e^{-is_n L}\right)\cosh(\gamma_n y)}{\gamma_n^4-\mu^4}, \label{d22}  
\end{align}
where the coefficients $a_j$, for $j=1\ldots 8$, are the unknown constants to be determined using the physical behavior of the vertical elastic plates at the edges. To that end, by adding and subtracting Eqs. \eqref{d21} and \eqref{d22}, we find out that 
\begin{align}
W^{\pm}(y)=& a^{\pm}_{1,5}\cos(\mu y)+a^{\pm}_{2,6}\sin(\mu y)+a^{\pm}_{3,7}\cosh(\mu y)+a^{\pm}_{4,8}\sinh(\mu y)
\nonumber
\\
&+2i\alpha \sum_{n=0}^{\infty} \dfrac{\chi_n^{\mp}\Pi_n^{\pm} \cosh(\gamma_{n}y)}{\gamma_{n}^{4}-\mu^{4}},
\label{d23} 
\end{align}
in terms of the symmetric and anti-symmetric modes, 
\begin{align*} 
W^{\pm}(y):= w^-(y)\pm w^+(y), 
\quad\text{where}\quad 
a^{\pm}_{j,j+4}:=a_j\pm a_{j+4},  \qquad j=1,2,3,4. 
\end{align*} 
Consequently, straightforward calculations indicate that $a_j$, for $j=1\ldots 8$, are obtained as 
\begin{align*}
a_j:=\dfrac{1}{2}\left(a^+_{j,j+4}+a^-_{j,j+4}\right)
\quad \text{and}\quad 
a_{j+4}:=\dfrac{1}{2}\left(a^+_{j,j+4}-a^-_{j,j+4}\right), \qquad j=1,2,3,4. 
\end{align*}

\subsection{Edge conditions on the vertical plates}

To ensure the uniqueness of a physical solution, we impose some edge conditions depending on the type of connection of vertical plates at finite edges $y=a$ and $y=b$. We entertain two different types of edge connections below. Specifically, we consider clamped and pin-joint connections. 

\subsubsection{Clamped edges of the vertical plates}

When the plates are clamped at the edges, both the displacements and their gradients are zero at the connection, i.e., 
\begin{align}
W^\pm(a)=0=W^\pm(b)
\quad\text{and}\quad
W_y^\pm(a)=0=W_y^\pm(b).
\label{d26}
\end{align}
The edge conditions \eqref{d26} and expression \eqref{d23} provide
\begin{align}
&a^{\pm}_{1,5}\cos(\mu \zeta)+a^{\pm}_{2,6}\sin(\mu \zeta)+a^{\pm}_{3,7}\cosh(\mu \zeta)+a^{\pm}_{4,8}\sinh(\mu \zeta)=-
2i\alpha\sum_{n=0}^\infty \frac{\chi^\mp\Pi^{\pm}_n \cosh(\gamma_n  \zeta)}{\gamma_n^4-\mu^4},
\label{527}
\\
&a^{\pm}_{1,5}\sin(\mu \zeta)-a^{\pm}_{2,6}\cos(\mu  \zeta)-a^{\pm}_{3,7}\sinh(\mu  \zeta)-a^{\pm}_{4,8}\cosh(\mu  \zeta)=
\dfrac{2i\alpha}{\mu}\sum_{n=0}^\infty\frac{\chi^\mp \Pi_m^\pm\gamma_n\sinh(\gamma_n  \zeta)}{\gamma_n^4-\mu^4},
\label{528}
\end{align}
for $\zeta=a$ or $b$.
The unknowns $a_{1,5}^\pm$, $a_{2,6}^\pm$, $a_{3,7}^\pm$, and $a_{4,8}^\pm$ can easily be recovered from Eqs. \eqref{527} and \eqref{528}.

\subsubsection{Pin-jointed edges of the vertical plates}

When the edges of the vertical elastic plates are pin-jointed, the displacements and the  bending moments of the plates are zero, i.e.,  
\begin{align}
W^\pm(a)=0=W^\pm(b)
\quad\text{and}\quad 
W_{yy}^\pm(a)=0=W_{yy}^\pm(b).\label{d29}
\end{align}
Imposing conditions \eqref{d29} on expression \eqref{d23} for $\zeta=a$ or $b$, we get
\begin{align}
&a^{\pm}_{1,5}\cos(\mu \zeta)+a^{\pm}_{2,6}\sin(\mu \zeta)+a^{\pm}_{3,7}\cosh(\mu \zeta)+a^{\pm}_{4,8}\sinh(\mu \zeta)=-
2i\alpha\sum_{n=0}^\infty \frac{\chi^\mp\Pi^{\pm}_n \cosh(\gamma_n  \zeta)}{\gamma_n^4-\mu^4},
\label{529}
\\
&a^{\pm}_{1,5}\cos(\mu \zeta)+a^{\pm}_{2,6}\sin(\mu  \zeta)-a^{\pm}_{3,7}\cosh(\mu  \zeta)-a^{\pm}_{4,8}\sinh(\mu  \zeta)=
\dfrac{2i\alpha}{\mu^2}\sum_{n=0}^\infty\frac{\chi^\mp \Pi_n^\pm\gamma_n^2\cosh(\gamma_n  \zeta)}{\gamma_n^4-\mu^4}.
\label{530}
\end{align}
Therefore, the unknowns $a^{\pm}_{1,5}$, $a^{\pm}_{2,6}$, $a_{3,7}^{\pm}$, and $a_{4,8}^{\pm}$ can be found from Eqs. \eqref{529} and \eqref{530}. 

\subsection{Continuity of velocity flux and horizontal edge conditions}

The  velocity  flux  conditions \eqref{eq:14aa} and \eqref{eq:14aaa} link the acoustic  velocity of the central region to the velocities of the inlet and outlet regions at $x=\pm  L$. These conditions lead to the amplitudes of symmetric and anti-symmetric modes propagating in the central region as 
\begin{align}
\chi^\pm_m=&\dfrac{i\Delta_{2,m} \left(V^\mp_1 +(\gamma^2_m+2)  V_{2}^\mp\right)}{2s_m\Pi^\pm_m}+\dfrac{i\alpha}{2s_m\Gamma_{2,m}\Pi^\pm_m} \left(F_\ell\xi_\ell R_{\ell,m}-\sum_{n=0}^{\infty}    \xi_n R_{n,m}  \Psi^\pm_n\right)
\nonumber
\\&+
\dfrac{\alpha}{2s_m\Gamma_{2,m}\Pi^\pm_m}\sum_{j=1}^4 a^\mp_{j,j+4} A_{j,m}
+\dfrac{i\alpha^2}{s_m\Gamma_{2,m} \Pi^\pm_m} \sum_{n=0}^\infty \frac{ \Pi^\mp_nT_{m,n}\chi^\pm_n }{\gamma_n^4-\mu^4},
\label{eq:23}       
\end{align}
where 
\begin{eqnarray*}
\begin{array}{ll}
A_{1,m}:=\ds\int_{a}^{b}\cos(\mu y)\cosh(\gamma_{m}y)dy,
&A_{2,m}:=\ds\int_{a}^{b}\sin(\mu y)\cosh(\gamma_{m}y)dy,
\\
A_{3,m}:=\ds\int_{a}^{b}\cosh(\mu y)\cosh(\gamma_{m}y)dy,
&A_{4,m}:=\ds\int_{a}^{b}\sinh(\mu y)\cosh(\gamma_{m}y)dy,
\end{array}
\end{eqnarray*}
and 
\begin{align}
T_{m,n}:=&\int_{a}^{b}\cosh(\gamma_{m}y)\cosh(\gamma_{n}y)dy, \quad
V_{1}^{\pm}:=e_{5}\pm  e_{7}, \quad 
V_{2}^{\pm}:=e_{6}\pm  e_{8},
\label{eq:47a}
\end{align}
The auxiliary unknowns $V_1^{\pm}$  and  $V_2^{\pm}$ involve additional constants,  
\begin{align}
e_{5}:=-i\phi_{2xyyy}(-L,b), \quad 
e_{6}:=-i\phi_{2xy}(-L,b), \quad 
e_{7}:=-i\phi_{2xyyy}(L,b), \quad 
e_{8}:=-i\phi_{2xy}(L,b),
\label{e58}
\end{align}
due to the normalization of velocity flux conditions \eqref{eq:14aa} and 
\eqref{eq:14aaa} with the generalized orthogonality relation \eqref{eq:8}. We refer the readers to Appendix \ref{append:a} for a detailed derivation of Eq. \eqref{eq:23}.

We are finally in possession of the required number of equations involving $\Psi^\pm_m$ and $\chi^\pm_m$, but we still need to cater for the auxiliary unknowns, $U_{j}^{\pm}$ and $V_{j}^{\pm}$, for $j=1,2$. They can be found by taking into account the physical behavior of horizontal elastic plates at the edges. To that end, we consider the situations when the horizontal elastic plates are both clamped and pin-jointed. The evaluation of the constants, $U_{j}^{\pm}$ and $V_{j}^{\pm}$ for both types of connections is discussed separately. It is worth pointing out that other physical conditions or combinations can also be imposed and that the framework is not limited only to the cases discussed below. However, for brevity and to avoid redundancy, we do not report them here. 

\subsubsection{Horizontal elastic plates with clamped edges}\label{subsec:3:3}

Setting the horizontal plate displacements as well as their gradients at the edges equal to zero, we obtain
\begin{eqnarray}
&\phi_{1y\phantom{x}}(-L,a)=0, &\phi_{3y\phantom{x}}(L,a)=0,
\label{eq:38}
\\
&\phi_{1xy}(-L,a)=0, &\phi_{3xy}(L,a)=0,
\label{eq:38a}
\\
&\phi_{2y\phantom{x}}(\pm  L,b)=0, &
\label{eq:39}
\\
&\phi_{2xy}(\pm L,b)=0. &
\label{eq:40}
\end{eqnarray}
Thanks to conditions \eqref{eq:38}, \eqref{eq:38a}, \eqref{eq:39}, and \eqref{eq:40}, we have $U_{2}^{\pm}=0= V_{2}^{\pm}$ and 
\begin{align}
U^\pm_{1}=&  \dfrac{2 }{S_1} \left(F_{\ell} \xi_{\ell} \Delta_{1,\ell} \Gamma_{1,\ell} +i\alpha  \sum_{m,n=0}^{\infty}\xi_{m} \Delta_{1,m} R_{m,n}\Pi^\mp_n \chi ^{\pm}_{n} \right),
\label{eq:28}
\\
V^+_1  =&  -\dfrac{i\alpha}{2 S_2}\sum_{m=0}^{\infty} \dfrac{\Delta_{2,m}\Pi_m^+}{s_m\Pi_m^-}  \left(F_\ell \xi_\ell R_{m,\ell}-\sum_{n=0}^{\infty}     \xi_n R_{m,n} \Psi^+_n\right)
\nonumber
\\
&-\dfrac{\alpha}{2S_2} \sum_{m=0}^{\infty} \left(\dfrac{\Delta_{2,m}\Pi_m^+}{s_m\Pi_m^-}\sum_{j=1}^4 a_{j,j+4}^- A_{j,m}\right)
\nonumber
\\
&- \dfrac{i\alpha^2}{S_2} \sum_{m,n=0}^{\infty} \dfrac{\Delta_{2,m}\Pi_m^+\Pi^+_n T_{m,n}  \chi^+_{n}}{s_m\Pi_m^-(\gamma^4_n-\mu^4)},
\label{eq:29q}
\\
V^-_{1} =&-\dfrac{i\alpha}{2 S_3}\sum_{m=0}^{\infty} \dfrac{\Delta_{2,m}\Pi_m^-}{s_m\Pi_m^+} \left(F_\ell \xi_\ell R_{m,\ell}-\sum_{n=0}^{\infty}   \xi_n R_{m,n}  \Psi^-_n\right)
\nonumber
\\
&-\dfrac{\alpha}{2S_3} \sum_{m=0}^{\infty}\left(\dfrac{\Delta_{2,m}\Pi_m^-}{s_m\Pi_m^+} \sum_{j=1}^4 a_{j,j+4}^+ A_{j,m}\right)
\nonumber 
\\
&- \dfrac{i\alpha^2}{S_3} \sum_{m,n=0}^{\infty}\dfrac{\Delta_{2,m}\Pi_m^- \Pi^-_n T_{m,n}  \chi^-_n}{s_m\Pi_n^+\left(\gamma^4_n-\mu^4\right)},
\label{eq:291q}
\end{align}
with 
\begin{align}
S_{1}:=\sum_{m=0}^{\infty} \Delta^2_{1,m} \Gamma_{1,m}\xi_{m},
\quad
S_{2}:=\sum_{m=0}^{\infty}\dfrac{i\Delta_{2,m}^2 \Gamma_{2,m} \Pi_m^+}{2s_m\Pi_m^-},
\quad
S_{3}:=\sum_{m=0}^{\infty}\dfrac{i\Delta_{2,m}^2 \Gamma_{2,m} \Pi_m^-}{2s_m\Pi_m^+}.\label{s1-3}
\end{align}
We refer the reader to Appendix \ref{appendix:b} for the derivation of expressions \eqref{eq:28},  \eqref{eq:29q}, and \eqref{eq:291q}.

\subsubsection{Horizontal elastic plates with pin-joint edges}\label{subsec3:4}

For  pin-joint horizontal edge  conditions, zero displacements and zero bending  moments of the horizontal plates lead to 
\begin{eqnarray}
\phi_{1y\phantom{xx}}(-L,a)=0, &&\phi_{3y\phantom{xx}}(L,a)=0, 
\label{eq:38b}
\\
\phi_{1yxx}(-L,a)=0, &&\phi_{3yxx}(L,a)=0,
\label{eq:39c}
\\
\phi_{2y\phantom{xx}}(-L,b)=0, &&\phi_{2y\phantom{xx}}(L,b)=0,
\label{eq:38d}
\\
\phi_{2yxx}(-L,b)=0, &&\phi_{2yxx}(L,b)=0.  
\label{eq:40e}
\end{eqnarray}
Note that Eq. \eqref{eq:38b} immediately renders $U^{\pm}_2=0$.  To  obtain  $  U^{\pm}_{1}$, we use Eqs. \eqref{eq:19} and \eqref{eq:39c}. We  multiply the equation for $\Psi^{\pm}_m$ with $\Delta_{1,m}\Gamma_{1,m}\xi^2_{m} $, take summation over $m$, and then utilize edge conditions \eqref{eq:38b}, after fairly easy manipulations, it is found that
\begin{align}
 U^\pm_1=&  \sum_{m,n=0}^{\infty}\dfrac{2i\alpha }{S_{4}}\Delta_{1,m} \xi_m^2 R_{m,n}\Pi^\mp_n \Psi^\pm_{n},
\label{eq:28ab1}
\end{align}
where 
\begin{align}
S_{4}:=\sum_{m=0}^{\infty}\Delta^2_{1,m} \xi^2_{m} \Gamma_{1,m}.\label{s4}
\end{align}
To determine $V^{\pm}_1$ and $V^{\pm}_2$, we multiply Eq. \eqref{eq:23} with $\Delta_{2,m}\Pi_m^{\pm}\Gamma_{2,m}$ and $\Delta_{2,m}\Pi_m^{\pm}\Gamma_{2,m} s^{2}_m$, take summation over $m$, and then use respective conditions \eqref{eq:38d} and \eqref{eq:40e},  we find after fairly easy manipulations that
\begin{align}
 S_{2}  V^+_{1}+  S_{5}  V^+_{2} =&  -i\alpha\sum_{m=0}^{\infty} \dfrac{\Delta_{2,m}\Pi_m^+}{2s_m\Pi_m^-}  \left(F_\ell \xi_\ell R_{m,\ell}-\sum_{n=0}^{\infty}     \xi_n R_{m,n} \Psi^+_n\right)
\nonumber
\\
&-\alpha \sum_{m=0}^{\infty} \left(\dfrac{\Delta_{2,m}\Pi_m^+}{2s_m\Pi_m^-}\sum_{j=1}^4 a_{j,j+4}^- A_{j,m}\right)
\nonumber
\\
&-i\alpha^{2}\sum_{m,n=0}^{\infty} \dfrac{\Delta_{2,m}\Pi_m^+\Pi^+_n T_{m,n}  \chi^+_{n}}{s_m\Pi_m^-(\gamma^4_n-\mu^4)},
\label{eq:29bbba}
\\
S_3  V^-_1+  S_6  V^-_2 =&- i\alpha\sum_{m=0}^{\infty} \dfrac{\Delta_{2,m}\Pi_m^-}{2s_m\Pi_m^+} \left(F_\ell \xi_\ell R_{m,\ell}-\sum_{n=0}^{\infty}   \xi_n R_{m,n}  \Psi^-_n\right)
\nonumber
\\
&-\alpha \sum_{m=0}^{\infty}\left(\dfrac{\Delta_{2,m}\Pi_m^-}{2s_m\Pi_m^+} \sum_{j=1}^4 a_{j,j+4}^+ A_{j,m}\right)
\nonumber 
\\
&- i\alpha^{2}\sum_{m,n=0}^{\infty}\dfrac{\Delta_{2,m}\Pi_m^- \Pi^-_n T_{m,n}  \chi^-_n}{s_m\Pi_n^+\left(\gamma^4_n-\mu^4\right)},
\label{eq:29bbb}
\end{align}
\begin{align}
 S_{7}  V^+_1+  S_{8}  V^+_2 =&  -i\alpha\sum_{m=0}^{\infty} \dfrac{\Delta_{2,m}s_m\Pi_m^+}{2\Pi_m^-}  \left(F_\ell \xi_\ell R_{m,\ell}-\sum_{n=0}^{\infty}     \xi_n R_{m,n} \Psi^+_n\right)
\nonumber
\\
&-\alpha \sum_{m=0}^{\infty} \left(\dfrac{\Delta_{2,m}s_m\Pi_m^+}{2\Pi_m^-}\sum_{j=1}^4 a_{j,j+4}^- A_{j,m}\right)
\nonumber
\\
&-i\alpha^{2}\sum_{m,n=0}^{\infty} \dfrac{\Delta_{2,m}s_m\Pi_m^+\Pi^+_n T_{m,n}  \chi^+_{n}}{\Pi_m^-(\gamma^4_n-\mu^4)},
\label{eq:29bbbd}
\\
 S_{9}  V^-_{1}+  S_{10}  V^-_{2} =&- i\alpha\sum_{m=0}^{\infty} \dfrac{\Delta_{2,m}s_m\Pi_m^-}{2\Pi_m^+} \left(F_\ell \xi_\ell R_{m,\ell}-\sum_{n=0}^{\infty}   \xi_n R_{m,n}  \Psi^-_n\right)
\nonumber
\\
&-\alpha \sum_{m=0}^{\infty}\left(\dfrac{\Delta_{2,m}s_m\Pi_m^-}{2\Pi_m^+} \sum_{j=1}^4 a_{j,j+4}^+ A_{j,m}\right)
\nonumber 
\\
&- i\alpha^{2}\sum_{m,n=0}^{\infty}\dfrac{\Delta_{2,m}s_m\Pi_m^- \Pi^-_n T_{m,n}  \chi^-_n}{\Pi_n^+\left(\gamma^4_n-\mu^4\right)},
\label{29bbbc}
\end{align}
where
\begin{eqnarray}
\left\{
\begin{array}{ll}
\ds S_{5}:=\sum_{m=0}^{\infty}\dfrac{i\Delta_{2,m}^2 \Gamma_{2,m} \Pi_m^+(\gamma^2_m+2)}{2s_m\Pi_m^-},
&\ds S_{6}:=\sum_{m=0}^{\infty}\dfrac{i\Delta_{2,m}^2 \Gamma_{2,m} \Pi_m^-(\gamma^2_m+2)}{2s_m\Pi_m^+},
\\
\ds S_{7}:=\sum_{m=0}^{\infty}\dfrac{i \Delta^2_{2,m} \Gamma_{2,m} s_{m}\Pi_m^+}{2\Pi_m^-},
&\ds S_{8}:=\sum_{m=0}^{\infty} \dfrac{i \Delta^2_{2,m} \Gamma_{2,m} s_{m}\Pi_m^+}{2\Pi_m^-}(\gamma^2_{m}+2),
\\
\ds S_{9}:=\sum_{m=0}^{\infty}  \dfrac{i\Delta_{2,m}^2\Gamma_{2,m}s_m\Pi_m^-}{2\Pi_m^+},
&\ds S_{10}:=\sum_{m=0}^{\infty} \dfrac{i\Delta_{2,m}^2\Gamma_{2,m}s_m\Pi_m^-}{2\Pi_m^+} (\gamma^2_{m}+2).
\end{array}
\right.
\label{s5-10}
\end{eqnarray}
The derivation of Eqs. \eqref{eq:28ab1}, \eqref{eq:29bbba}, \eqref{eq:29bbb}, \eqref{eq:29bbbd}, and \eqref{29bbbc} follows the same procedure as described in Appendix \ref{appendix:b} for the derivation of Eqs.  \eqref{eq:28}, \eqref{eq:29q}, and \eqref{eq:291q} with appropriate changes.

In a nutshell, we have derived the required number of equations for $\Psi_n^\pm$ and $\chi^\pm_n$ from the continuity of the pressure and velocity fluxes, for coefficients $a_j$'s from the edge conditions on the vertical elastic plates, and for the auxiliary unknowns $U_j^\pm$ and $V_j^\pm$ from the edge conditions on the horizontal elastic plates. The MMT - tailored-Galerkin solution of the scattering problems is obtained by solving the aforementioned systems of linear equations simultaneously.

\section{Modal approach}\label{sec:Modal}

In this section, we supplement the MMT with the modal approach to find the vertical elastic plate displacements $w^{\pm}$. The modal approach is useful as it broadens the range of edge conditions that can be addressed and avoids the need for additional root-finding. In this study, we will use the modal approach for comparing the results obtained using the tailored-Galerkin approach. In this approach, the vertical plate displacements are expressed as linear combinations of a set of known basis functions and the corresponding coefficients are sought. To that end, it is convenient to express the plate displacements in terms of the eigenfunctions of the duct height $y=b$. The properties of the eigenfunctions such as linear dependence and the identity \eqref{Gf} are critical for the success of the modal approach. The details beyond those presented in this section can be found in \cite{LawrieAfzal, Kim}. 

In the model approach, we express the vertical elastic plate displacements as 
\begin{align}
w^\pm(y)=&\sum_{n=0}^\infty G_{n}^\pm \cosh(\gamma_n y), \qquad y\in(a,b),
\label{eq441}
\end{align}
and determine the unknown modal coefficients, $G_{n}^\pm$, using prescribed boundary and edge conditions. To that end, we substitute Eq. \eqref{eq441} into Eq.  \eqref{d19} for $x=-L$, multiply the resultant equation by $\cosh(\gamma_m y)$, and integrate over  $(a,b)$ to get
\begin{align}
\sum_{n=0}^\infty G_n^-\left(\gamma_n^4-\mu^4\right)T_{n,m}=-\alpha\sum_{n=0}^\infty(B_n e^{-is_n L}+C_n e^{is_n L})T_{n,m}, 
\label{eq491}
\end{align}
where $T_{n,m}$ is the integral defined in Eq. \eqref{eq:47a}. Integration by parts suggests that 
\begin{align}
\gamma_n^2 T_{n,m}=& \gamma_m^2T_{n,m}+\gamma_n\cosh(\gamma_m b)\sinh(\gamma_n b)-\gamma_n\cosh(\gamma_m a)\sinh(\gamma_n a)
\nonumber \\
&+\gamma_m\cosh(\gamma_n a)\sinh(\gamma_m a)-\gamma_m\cosh(\gamma_n b)\sinh(\gamma_m b).
\label{521}  
\end{align}
Moreover, after fairly simple manipulations, Eq. \eqref{521} renders the identity
\begin{align}
\label{d38}
\gamma_n^4 T_{n,m}=& \gamma_m^4T_{n,m}+\gamma_m^2\gamma_n\cosh(\gamma_m b)\sinh(\gamma_n b)-\gamma_m^2\gamma_n\cosh(\gamma_m a)\sinh(\gamma_n a)
\nonumber \\
&+\gamma_m^3\cosh(\gamma_n a)\sinh(\gamma_m a)-\gamma_m^3\cosh(\gamma_n b)\sinh(\gamma_m b)
\nonumber \\
&+\gamma_n^3\cosh(\gamma_m b)\sinh(\gamma_n b)-\gamma_n^3\cosh(\gamma_m a)\sinh(\gamma_n a)
\nonumber \\
&+\gamma_m\gamma_n^2\cosh(\gamma_n a)\sinh(\gamma_m a)-\gamma_m\gamma_n^2\cosh(\gamma_n b)\sinh(\gamma_m b).
\end{align}
Substituting Eq. \eqref{d38} into Eq. \eqref{eq491}, it is found that 
\begin{align}
\label{601}
 \sum_{n=0}^\infty G_n^-T_{n,m}=&-\frac{L_1\cosh(\gamma_m a)}{\gamma_m^4-\mu^4}-\frac{L_2\cosh(\gamma_m b)}{\gamma_m^4-\mu^4}-\frac{L_3\gamma_m\sinh(\gamma_m a)}{\gamma_m^4-\mu^4}-\frac{L_4\gamma_m\sinh(\gamma_m b)}{\gamma_m^4-\mu^4}
\nonumber\\
&-\frac{L_5\gamma_m^2\cosh(\gamma_m a)}{\gamma_m^4-\mu^4}-\frac{L_6\gamma_m^2\cosh(\gamma_m b)}{\gamma_m^4-\mu^4}-\frac{L_7\gamma_m^3\sinh(\gamma_m a)}{\gamma_m^4-\mu^4}
\nonumber\\
&-\frac{L_8\gamma_m^3\sinh(\gamma_m b)}{\gamma_m^4-\mu^4} -\frac{\alpha}{\gamma_m^4-\mu^4}\sum_{n=0}^\infty(B_n e^{-is_n L}+C_n e^{is_n L})T_{n,m},
\end{align}
where the auxiliary constants, $L_j$, for $j=1,\cdots, 8$, are defined as 
\begin{eqnarray*} 
\begin{array}{llll}
L_1:=-w^-_{yyy}(a), &L_2:=w^-_{yyy}(b), &L_3:=w^-_{yy}(a), &L_4:=-w^-_{yy}(b),
\\
L_5:=-w^-_{y}(a), &L_6:=w^-_{y}(b), &L_7:=w^-(a), &L_8:=-w^-(b).
\end{array}
\end{eqnarray*} 

On multiplying  Eq. \ref{601} by $\alpha\cosh(\gamma_m y)/\Gamma_{2,m}$ and summing over index $m$, we get 
\begin{align}
\alpha\sum_{n,m=0}^\infty\frac{G_n^-\cosh(\gamma_m y)T_{n,m}}{\Gamma_{2,m}}=
&-\sum_{j=1}^8 L_j\varphi^j(y)-\alpha^2\sum_{n,m=0}^\infty\frac{(B_n e^{-is_n L}+C_n e^{is_n L})\cosh(\gamma_m y)T_{n,m}}{\Gamma_{2,m}(\gamma_m^4-\mu^4)},
\label{611}  
\end{align}
where the functions $\varphi^j(y)$, for  $j=1, \cdots, 8$, are defined as
\begin{eqnarray*}
\begin{array}{llll}
\varphi^1(y):=z(y,a), &\varphi^2(y):=z(y,b), &\varphi^3(y):=z'(y, a), &\varphi^4(y):=z'(y, b),
\\
\varphi^5(y):=z''(y, a), &\varphi^6(y):=z''(y, b),&\varphi^7(y):=z'''(y, a), &\varphi^8(y):=z'''(y, b).
\end{array}
\end{eqnarray*} 
Here, 
\begin{equation*}
z(y, \zeta)=\alpha\sum_{m=0}^\infty\frac{\cosh(\gamma_m y)\cosh(\gamma_m \zeta)}{\Gamma_{2,m}(\gamma_m^4-\mu^4)}, \qquad \zeta = a, b,
\end{equation*}
and the superposed prime indicates the derivative with respect to $\zeta$.

Recall that our aim is to recover coefficients $G_{n}^-$ for constructing $w^-(y)$ from  Eq. \eqref{611}. Unfortunately, no explicit orthogonality relation exists for the functions $\cosh(\gamma_n y)$ over $[a,b]$. To that end, we will make use of the identity \eqref{Gf} in lieu.  On interchanging the orders of the summation and integration on the left hand side of Eq. \eqref{611} and invoking  identity \eqref{Gf}, we get
\begin{align}
w^-(y)=&-\sum_{j=1}^8 L_j\varphi^j(y)
-\alpha^2\sum_{n,m=0}^\infty\frac{(B_n e^{-is_n L}+C_n e^{is_n L})\cosh(\gamma_my)T_{n,m}}{\Gamma_{2,m}(\gamma_m^4-\mu^4)}.
\label{5311a}  
\end{align}

Similarly, adopting the same procedure for $w^+(y)$, we find out that  
\begin{align}
w^+(y)=&-\sum_{j=1}^8 M_j\varphi^j(y)
+\alpha^2\sum_{n,m=0}^\infty\frac{(B_n e^{is_n L}+C_n e^{-is_n L})\cosh(\gamma_my)T_{n,m}}{\Gamma_{2,m}(\gamma_m^4-\mu^4)},
\label{5312} 
\end{align} 
where auxiliary constants, $M_j$, for $j=1,\cdots, 8$, are defined as
\begin{eqnarray*} 
\begin{array}{llll}
M_1:=-w^+_{yyy}(a), &M_2:=w^+_{yyy}(b), &M_3:=w^+_{yy}(a), & M_4:=-w^+_{yy}(b),
\\
M_5:=-w^+_{y}(a), &M_6:=w^+_{y}(b), &M_7:=w^+(a), &M_8:=-w^+(b).
\end{array}
\end{eqnarray*}
Finally, we add and subtract Eqs. \eqref{5311a} and  \eqref{5312} to get
 \begin{align}
 W^\pm(y)=&-\sum_{j=1}^8 N^\pm_j\varphi^j(y)+2i\alpha^2 \sum_{n,m=0}^\infty \dfrac{\chi^\mp_n \Pi_n^\pm \cosh(\gamma_m y) T_{n,m}}{\Gamma_{2,m}(\gamma_q^4-\mu^4)},
 \label{541}
\end{align}
where $N_j^{\pm}:= L_j \pm M_j$, for $j=1,\ldots, 8$. 

It should be noted that the displacements of the vertical elastic plates, $W^\pm$, still involve auxiliary unknowns, $N^\pm_j$. These constants can be determined by imposing physical conditions on the edge connections of the vertical elastic plates, as in the tailored-Galerkin approach. After determining $N_j^\pm$, it is simple to compute $L_j$ and $M_j$ as
\begin{align*}
L_j= \frac{N_j^++N_j^-}{2} \quad \text{and}\quad M_j:= \frac{N_j^+-N_j^-}{2},\qquad j=1,\cdots, 8.
\end{align*}

We discuss the recovery $N^\pm_j$ in Section \ref{ss:ModalVer}. To derive the required number of equations for the linear system of the MMT using the modal expansions of the vertical displacements, we use the continuity of the velocity flux and the physical conditions on the edge connections of the horizontal elastic plates in Section \ref{ss:ModalCVF}. 

\subsection{Edge conditions on the vertical plates}\label{ss:ModalVer}
  
In this section, we assume either clamped or pin-joint edge connections of both the vertical elastic plates to determine $N^\pm_j$. Individual cases are discussed separately. As mentioned for the tailored-Galerkin approach, other physically relevant edge conditions can also be imposed, and the framework is not limited only to both clamped and both pin-joint edge connections.

\subsubsection{Clamped edges of the vertical plates}\label{subsec:4:4}

For  clamped edges of the vertical plates at $x=\pm L$, Eq. \eqref{d26} for zero plate displacements and zero gradients, together with Eq. \eqref{541},  furnishes
\begin{equation*}
N_5^{\pm}=N_6^{\pm} =N_7^{\pm}=N_8^{\pm}=0,
\end{equation*}
and, for $\zeta=a$ or $b$, 
\begin{align}
\sum_{j=1}^4 N_j^\pm\varphi^j(\zeta)=& 2i\alpha^2\sum_{n,m=0}^\infty\frac{\chi^\mp\Pi_n^\mp \cosh(\gamma_m  \zeta)T_{n,m}}{\Gamma_{2,m}(\gamma_m^4-\mu^4)},
\label{571}
\\
\sum_{j=1}^4 N_j^\pm\varphi^j_y(\zeta)=&
2i\alpha^2\sum_{n,m=0}^\infty\frac{\chi^\mp \Pi_m^\mp\gamma_m\sinh(\gamma_m \zeta)T_{n,m}}{\Gamma_{2,m}(\gamma_m^4-\mu^4)}.
\label{561}
\end{align}
Solving Eqs. \eqref{571} and \eqref{561} simultaneously, one can easily find $N_j^\pm$, for $j=1,\cdots,4$. 
 
\subsubsection{Pin-jointed edges of the vertical plates}\label{subsec  4:5}

For  pin-joint edge connections of the vertical plates at $x=\pm L$, Eq. \eqref{d29} for zero plate displacements and zero bending moments, together with Eq.  \eqref{541}, leads to
\begin{equation*}
N_3^{\pm}=N_4^{\pm}=N_7^{\pm}=N_8^{\pm}=0,
\end{equation*}
and 
\begin{align}
N_1^\pm\varphi^1( \zeta)+N_2^\pm\varphi^2(\zeta)+N_5^\pm\varphi^5(\zeta)+N_6^\pm\varphi^6(\zeta)=&
2i\alpha^2\sum_{n,m=0}^\infty\frac{\chi_n^\mp\Pi_n^\pm\cosh(\gamma_m \zeta)T_{n,m}}{\Gamma_{2,m}(\gamma_m^4-\mu^4)},
\label{570}
\\
N^\pm_1\varphi^1_{yy}(\zeta)+N^\pm_2\varphi^2_{yy}(\zeta)+N^\pm_5\varphi^5_{yy}(p)+N^\pm_6\varphi^6_{yy}(\zeta)=&
2i\alpha^2\sum_{n,m=0}^\infty\frac{\chi_n^\mp \Pi_n^\pm\gamma_m^{2}\cosh(\gamma_m \zeta)T_{n,m}}{\Gamma_{2,m}(\gamma_m^4-\mu^4)}.
\label{5611}
\end{align}
Solving Eqs. \eqref{570} and \eqref{5611} simultaneously, one can find all the non-zero constants $N_j^\pm$, for $j=1,2,5,6$.

\subsection{Continuity of velocity flux and horizontal edge conditions}\label{ss:ModalCVF}
 
In order to complete the linear system of equations for MMT supplemented by the modal approach, we use the continuity of the velocity flux. To that end, Eqs.  \eqref{eq:14aa} and \eqref{eq:14aaa} lead to 
\begin{align}
\chi^\pm_m=&\dfrac{i\Delta_{2,m} \left(V^\mp_1 +(\gamma^2_m+2)  V_{2}^\mp\right)}{2s_m\Pi^\pm_m}+\dfrac{i\alpha}{2s_m\Gamma_{2,m}\Pi^\pm_m} \left(F_\ell\xi_\ell R_{\ell,m}-\sum_{n=0}^{\infty}    \xi_n R_{n,m}  \Psi^\pm_n\right)
\nonumber
\\&-
\dfrac{\alpha}{2s_m\Gamma_{2,m}\Pi^\pm_m}\sum_{j=1}^8 N^{\mp}_j \Phi^j_m
+\dfrac{i\alpha^2}{s_m\Gamma_{2,m} \Pi^\pm_m} \sum_{n,q=0}^\infty \frac{ \Pi^\mp_nT_{m,q}T_{n,q}\chi^\pm_n }{\gamma_n^4-\mu^4},
\label{eq:23M}       
\end{align}
where
\begin{align}
\Phi^j_m=\int_a^b \varphi^j(y)Y_2(\gamma_m,y)dy, \qquad j=1,\ldots,8.
\end{align}
We refer the reader to Appendix \ref{appendix:c} for the derivation of Eq. \eqref{eq:23M}.

It should be noted that we still have to recover the auxiliary unknowns $U_j^\pm$ and $V_j^\pm$, for $j=1,2$. To that end, different physical conditions on the edge connections of the horizontal elastic plates can be used. Below, we discuss the cases when both the horizontal elastic plates are either clamped or pin-jointed. 

\subsubsection{Horizontal elastic plates with clamped edges}\label{subsec:3:5}

For clamped edge conditions on horizontal plates, the quantities $U_1^{\pm}$ and $U_2^{\pm}$ are linked to the edges of elastic plates that have been defined in terms of field potentials of inlet and outlet regions, and their values remain the same as in Section \ref{subsec:3:3}. However, the auxiliary parameters $V_1^{\pm}$ and $V_2^{\pm}$ depend on the field potential of the chamber region and are different. To that end, we apply the conditions \eqref{eq:39} and \eqref{eq:40} together with the field potentials \eqref{eq:23M}. Accordingly, 
\begin{equation*}
V_2^{\pm}=e_2\pm e_4=0,
\end{equation*}
and
\begin{align}
V^+_1=&  -\dfrac{i\alpha}{2 S_2}\sum_{m=0}^{\infty} \dfrac{\Delta_{2,m}\Pi_m^+}{s_m\Pi_m^-}  \left(F_\ell \xi_\ell R_{m,\ell}-\sum_{n=0}^{\infty}     \xi_n R_{m,n} \Psi^+_n\right)
\nonumber
\\
&+\dfrac{\alpha}{2S_2} \sum_{m=0}^{\infty} \left(\dfrac{\Delta_{2,m}\Pi_m^+}{s_m\Pi_m^-}\sum_{j=1}^8 N^{-}_j \Phi^j_m\right)- \dfrac{i\alpha^2}{S_2} \sum_{m,n,q=0}^{\infty} \dfrac{\Delta_{2,m}\Pi_m^+\Pi^+_n T_{m,q}T_{n,q}  \chi^+_{n}}{s_m\Pi_m^-(\gamma^4_n-\mu^4)},
\label{611a}  
\\
V^-_1 =& -\dfrac{i\alpha}{2 S_3}\sum_{m=0}^{\infty} \dfrac{\Delta_{2,m}\Pi_m^-}{s_m\Pi_m^+} \left(F_\ell \xi_\ell R_{m,\ell}-\sum_{n=0}^{\infty}   \xi_n R_{m,n}  \Psi^-_n\right)
\nonumber
\\
&+\dfrac{\alpha}{2S_3} \sum_{m=0}^{\infty}\left(\dfrac{\Delta_{2,m}\Pi_m^-}{s_m\Pi_m^+} \sum_{j=1}^8 N^{+}_j \Phi^j_m\right)-\dfrac{i\alpha^2}{S_3} \sum_{m,n,q=0}^{\infty}\dfrac{\Delta_{2,m}\Pi_m^- \Pi^-_n T_{m,q}T_{n,q}   \chi^-_n}{s_m\Pi_n^+\left(\gamma^4_n-\mu^4\right)},
\label{611b}  
\end{align}
where $S_2$ and $S_3$ are defined in Eq. \eqref{s1-3}.

\subsubsection{Horizontal elastic plates with pin-joint edges}\label{subsec:3:6}

For pin-joint connections of the horizontal elastic plates, $U^\pm_2=0$  and  the values of $U_1^\pm$ are the same as in Eq. \eqref{eq:28ab1} derived for the tailored-Galerkin approach in Section \ref{subsec3:4}. To  determine $V^\pm_1$ and $V^\pm_2$,  we  apply conditions \eqref{eq:38d} and \eqref{eq:40e} together with  the  field  potentials  \eqref{eq:23M}. Therefore, we get
\begin{align}
 S_{2}  V^+_{1}+  S_{5}  V^+_{2} =&  -i\alpha\sum_{m=0}^{\infty} \dfrac{\Delta_{2,m}\Pi_m^+}{2s_m\Pi_m^-}  \left(F_\ell \xi_\ell R_{m,\ell}-\sum_{n=0}^{\infty}     \xi_n R_{m,n} \Psi^+_n\right)
\nonumber
\\
&+\alpha \sum_{m=0}^{\infty} \left(\dfrac{\Delta_{2,m}\Pi_m^+}{2s_m\Pi_m^-}\sum_{j=1}^8 N^{+}_j \Phi^j_m\right)-i\alpha^{2}\sum_{m,n,q=0}^{\infty} \dfrac{\Delta_{2,m}\Pi_m^+\Pi^+_n T_{m,q}T_{n,q}   \chi^+_{n}}{s_m\Pi_m^-(\gamma^4_n-\mu^4)},
\label{eq:29bb3}
\end{align}
\begin{align}
S_3  V^-_1+  S_6  V^-_2 =&- i\alpha\sum_{m=0}^{\infty} \dfrac{\Delta_{2,m}\Pi_m^-}{2s_m\Pi_m^+} \left(F_\ell \xi_\ell R_{m,\ell}-\sum_{n=0}^{\infty}   \xi_n R_{m,n}  \Psi^-_n\right)
\nonumber
\\
&+\alpha \sum_{m=0}^{\infty}\left(\dfrac{\Delta_{2,m}\Pi_m^-}{2s_m\Pi_m^+} \sum_{j=1}^8 N^{-}_j \Phi^j_m \right)- i\alpha^{2}\sum_{m,n,q=0}^{\infty}\dfrac{\Delta_{2,m}\Pi_m^- \Pi^-_n T_{m,q}T_{n,q}   \chi^-_n}{s_m\Pi_n^+\left(\gamma^4_n-\mu^4\right)},
\label{eq:29bb2}
\end{align}
\begin{align}
 S_{7}  V^+_1+  S_{8}  V^+_2 =&  -i\alpha\sum_{m=0}^{\infty} \dfrac{\Delta_{2,m}s_m\Pi_m^+}{2\Pi_m^-}  \left(F_\ell \xi_\ell R_{m,\ell}-\sum_{n=0}^{\infty}\xi_n R_{m,n} \Psi^+_n\right)
\nonumber
\\
&+\alpha \sum_{m=0}^{\infty} \left(\dfrac{\Delta_{2,m}s_m\Pi_m^+}{2\Pi_m^-}\sum_{j=1}^8 N^{+}_j \Phi^j_m\right)-i\alpha^{2}\sum_{m,n,q=0}^{\infty} \dfrac{\Delta_{2,m}s_m\Pi_m^+\Pi^+_n T_{m,q}T_{n,q}   \chi^+_{n}}{\Pi_m^-(\gamma^4_n-\mu^4)},
\label{29bb1}
\end{align}
\begin{align}
S_{9}  V^-_{1}+  S_{10}  V^-_{2} =&- i\alpha\sum_{m=0}^{\infty} \dfrac{\Delta_{2,m}s_m\Pi_m^-}{2\Pi_m^+} \left(F_\ell \xi_\ell R_{m,\ell}-\sum_{n=0}^{\infty}   \xi_n R_{m,n}  \Psi^-_n\right)
\nonumber
\\
&+\alpha \sum_{m=0}^{\infty}\left(\dfrac{\Delta_{2,m}s_m\Pi_m^-}{2\Pi_m^+} \sum_{j=1}^8 N^{-}_j \Phi^j_m\right)- i\alpha^{2}\sum_{m,n,q=0}^{\infty}\dfrac{\Delta_{2,m}s_m\Pi_m^- \Pi^-_n T_{m,q}T_{n,q}  \chi^-_n}{\Pi_n^+\left(\gamma^4_n-\mu^4\right)}.
\label{eq:29bb}
\end{align}
Here the constants $S_j$, for $j=4,\cdots,10$, are defined in Eqs. \eqref{s4} and \eqref{s5-10}.

\section{Numerical results and discussions}\label{sec:Num}

In this section, we first validate the mode-matching-tailored Galerkin approach (MTG) numerically. Then, we investigate the influence of the edge conditions on the reflected power, transmitted power, and transmission loss in different frequency bands and compare the results produced by the MTG with those of the mode-matching-modal approach (MMA). To that end, we truncate the infinite linear algebraic systems derived in Sections \ref{sec:TG} and \ref{sec:Modal} and retain the first $N$ terms for a sufficiently large $N\in\mathbb{Z}^+$. 

The numerical experiments are carried out on elastic plates (assumed to be of aluminum) with a thickness of $\bar{h}=0.0006$m and a density of $\rho_p=2700$kg/m$^3$. Following \cite{RN13Jasa}, the Young's modulus, Poisson's ratio, speed of sound, and density of air are fixed at $E=7.2\times 10^{10}$N/m$^2$, $\upsilon=0.34$, $c=344$m/s, and $\rho_a=1.2$kg/m$^3$, respectively. The dimensional height parameters are taken to be $\bar{a}=0.06$m and $\bar{b}=0.085$m. Two different incident fields, a structure-borne fundamental mode ($\ell=0$) and a fluid-borne second mode ($\ell=1$), are considered for the experiments. 

\subsection{Validation of the MTG}

\begin{figure}[htb!]
\subfigure[Real parts of the velocities.]
{\includegraphics[width=0.48\textwidth]{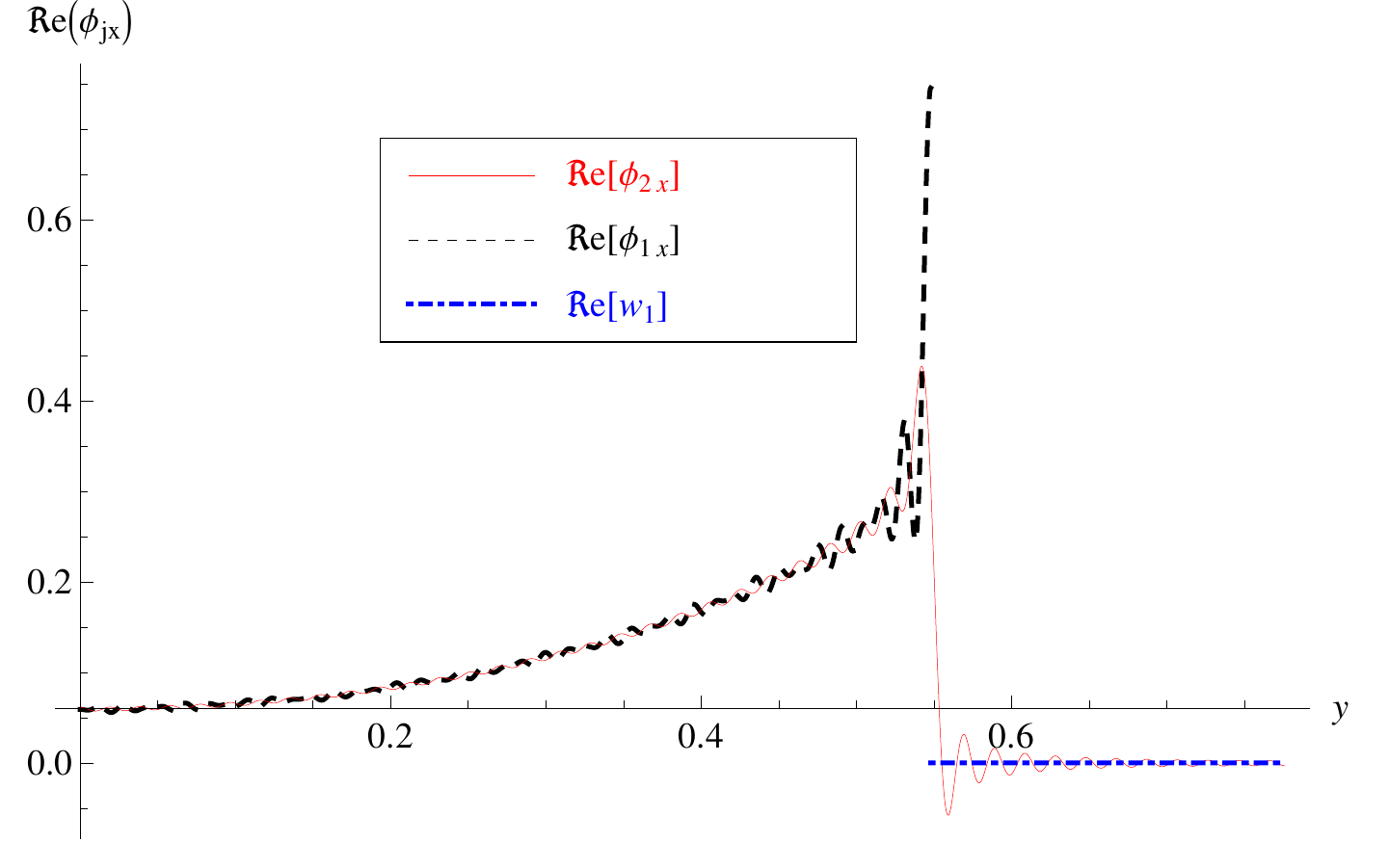}\label{fig:2a}}
\subfigure[Real parts of the pressures.]{\includegraphics[width=0.48\textwidth]{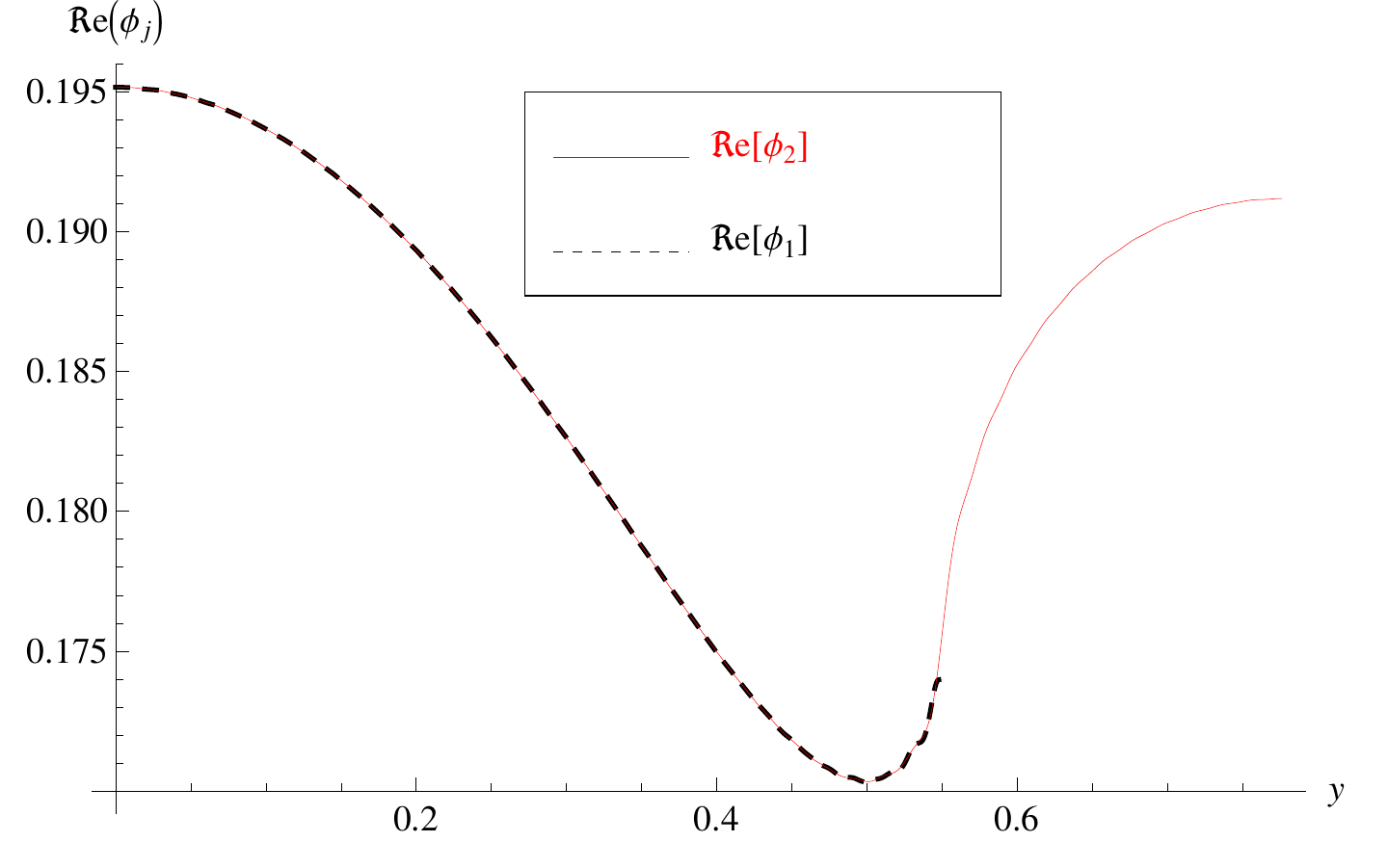}\label{fig:2b}}  
\subfigure[Imaginary parts of the velocities.]{\includegraphics[width=0.48\textwidth]{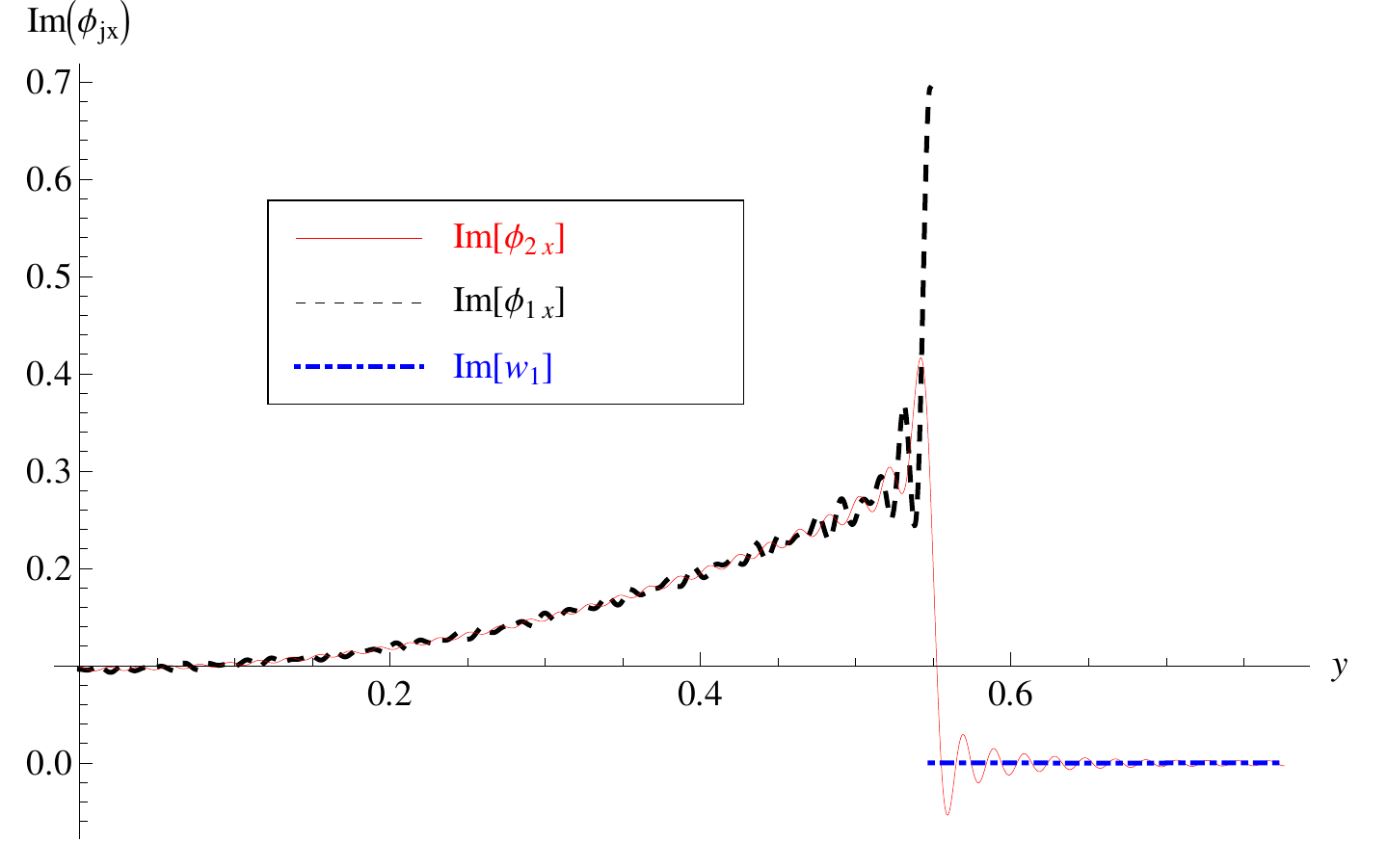}\label{fig:2c}}
\subfigure[Imaginary parts of the pressures.]{\includegraphics[width=0.48\textwidth]{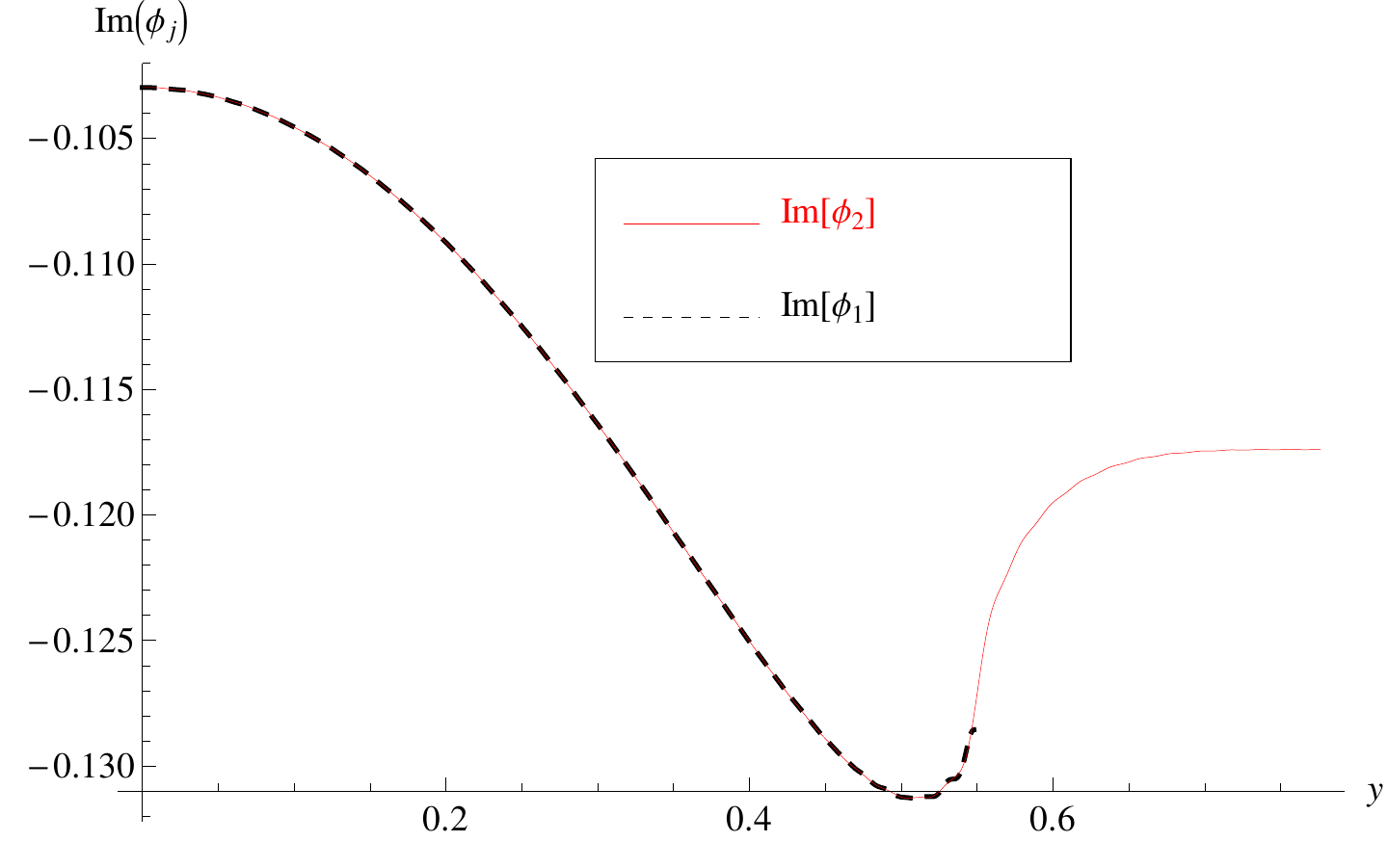}\label{fig:2d}}
\caption{The real and imaginary parts of the normal velocities and pressures versus height at the interface $\bar{x}=-\bar{L}$ in the presence of vertical elastic plates for clamped edges using frequency $f=700$Hz, $\overline{L}=0.045$m, and $N=80$.}
\label{fig:2}
\end{figure}

The MTG is validated by reconstructing the matching conditions using the truncated form of the MTG solution. In Fig. \ref{fig:2}, the real and imaginary parts of the non-dimensional normal velocities and pressures are plotted at interface $x=-L$. Simlar plots can be produced for the interface $x=-L$. The curves for real as well as imaginary parts of the normal velocities and pressures coincide at the interfaces in the range $0\leq y\leq b$. We conclude that the matching conditions \eqref{eq:21a} and \eqref{eq:14aa} are fully satisfied. Note that the curves oscillate around their mean values near the corners $y=a$ and $y=b$, which is the so-called Gibb's phenomenon, created by the truncation of a poorly convergent series \cite{RN13Jasa}. However, as $y$ moves away from the corners $y=a$ and $y=b$, the amplitude of the oscillation reduces significantly. 

\subsection{Scattering analysis}

In this section, we perform a power and transmission loss analysis to understand the scattering of fundamental and fluid-borne mode incident fields in the waveguide structure with a wave-bearing cavity and flexible inlet/outlet.

The reflected energy flux or power in the inlet and the transmitted energy flux or power in the outlet are defined as
\begin{align*}
{
\mathcal{E}_r:=\dfrac{1}{\alpha}\sum_{m=0}^{K-1} |A_m|^2 \xi_{m} \Gamma_{1,m}
%\label{ener1}
\quad\text{and}\quad
\mathcal{E}_t:=\dfrac{1}{\alpha}\sum_{m=0}^{K-1} |D_m|^2 \xi_{m} \Gamma_{1,m}, 
%\label{ener2}
}
\end{align*}
respectively; see, for instance, \cite{Warren}. The incident energy flux or power is normalized in this equation, and $K$ denotes the number of cut-on modes in the extended inlet/outlet region. The conservation of energy flux in a confined system is expressed through the conservation of power identity, 
\begin{equation*}
\mathcal{E}_r+\mathcal{E}_t=1,%\label{consL}
\end{equation*}
wherein the sum of the reflected and transmitted powers represents the total incident power fed to the system. The transmission loss of the transmitted power through the waveguide is defined as 
\begin{eqnarray}
{\rm TL}:=-10\log_{10}\left(\dfrac{\mathcal{E}_t}{\mathcal{E}_i}\right),
  \label{681}
\end{eqnarray}
where $\mathcal{E}_i$ is the incident power (being  normalized to $1$ herein); see, for instance, \cite{lawrie2006tuning}. Throughout this section, we fix $\overline{L}=0.02$m.

\subsubsection{Reflected and transmitted powers}

\begin{figure}[htb!]
	\subfigure[Same edge conditions with $N=40$, $\ell=0$.]{\includegraphics[width=.48\textwidth]{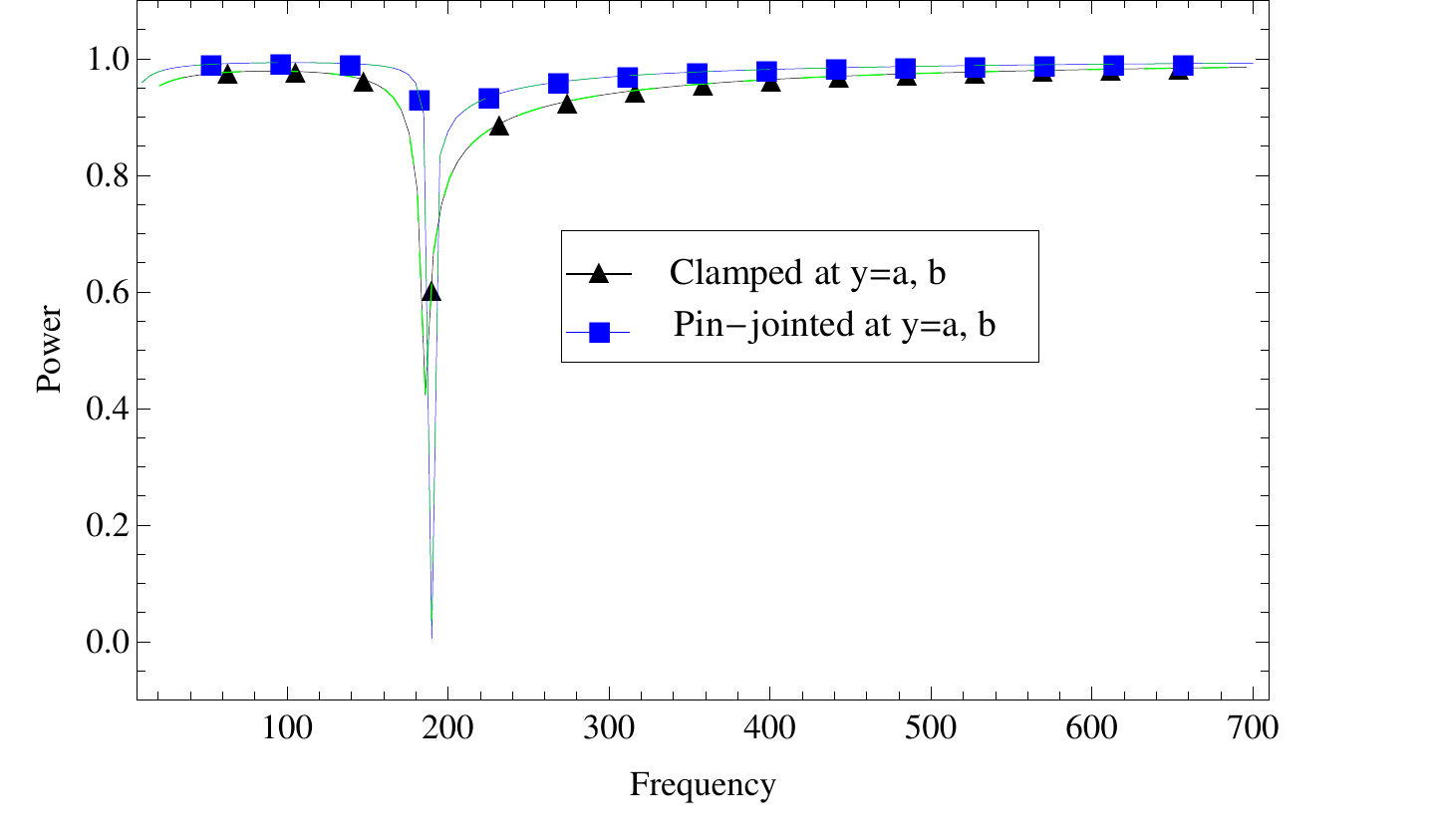}\label{fig:3a}}  
	\subfigure[Same edge conditions with $N=40$, $\ell=1$.]{\includegraphics[width=.48\textwidth]{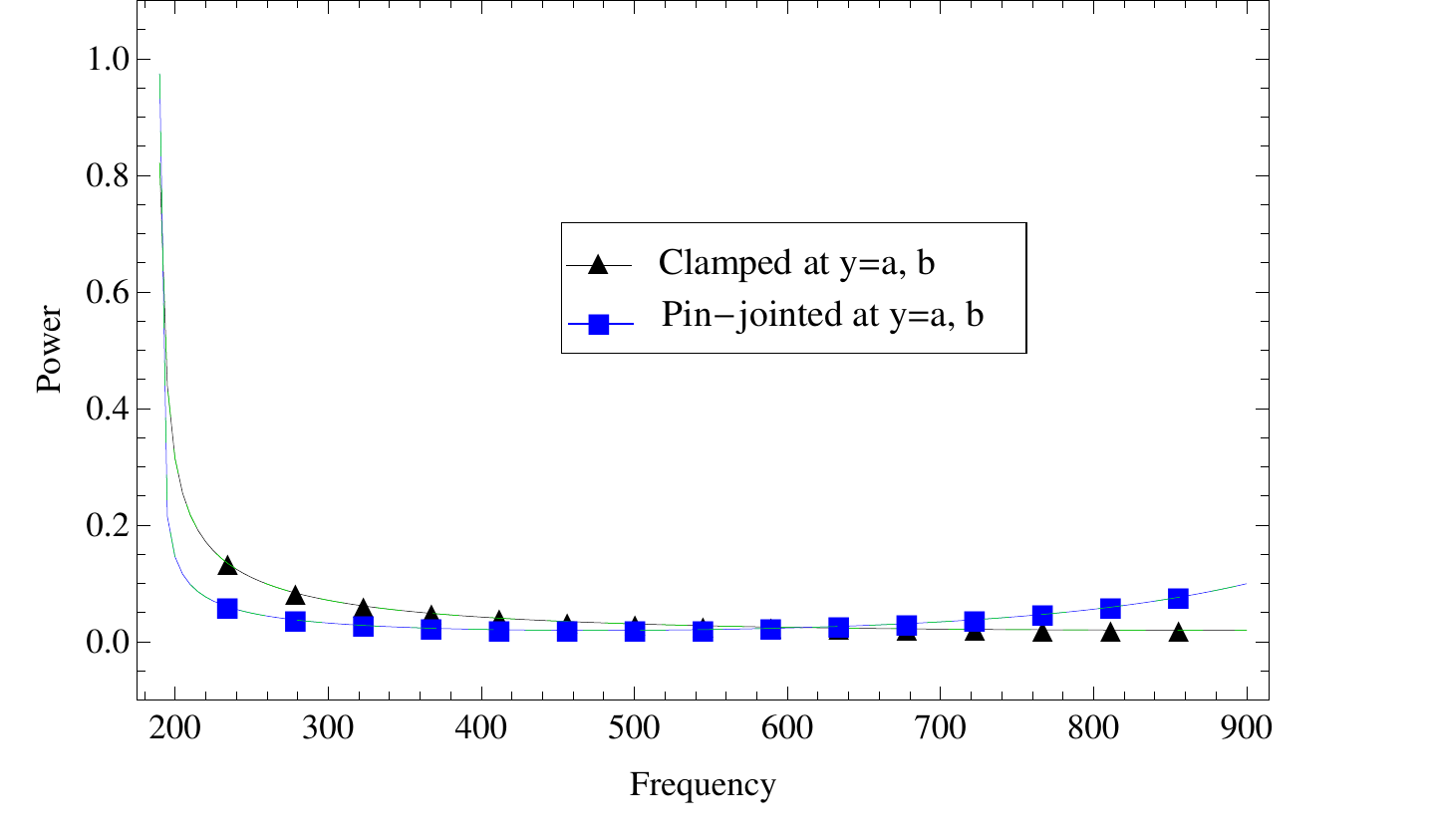}\label{fig:3b}}
\\
  	\subfigure[Mixed edge conditions with $N=25$, $\ell=0$.]{\includegraphics[width=.48\textwidth]{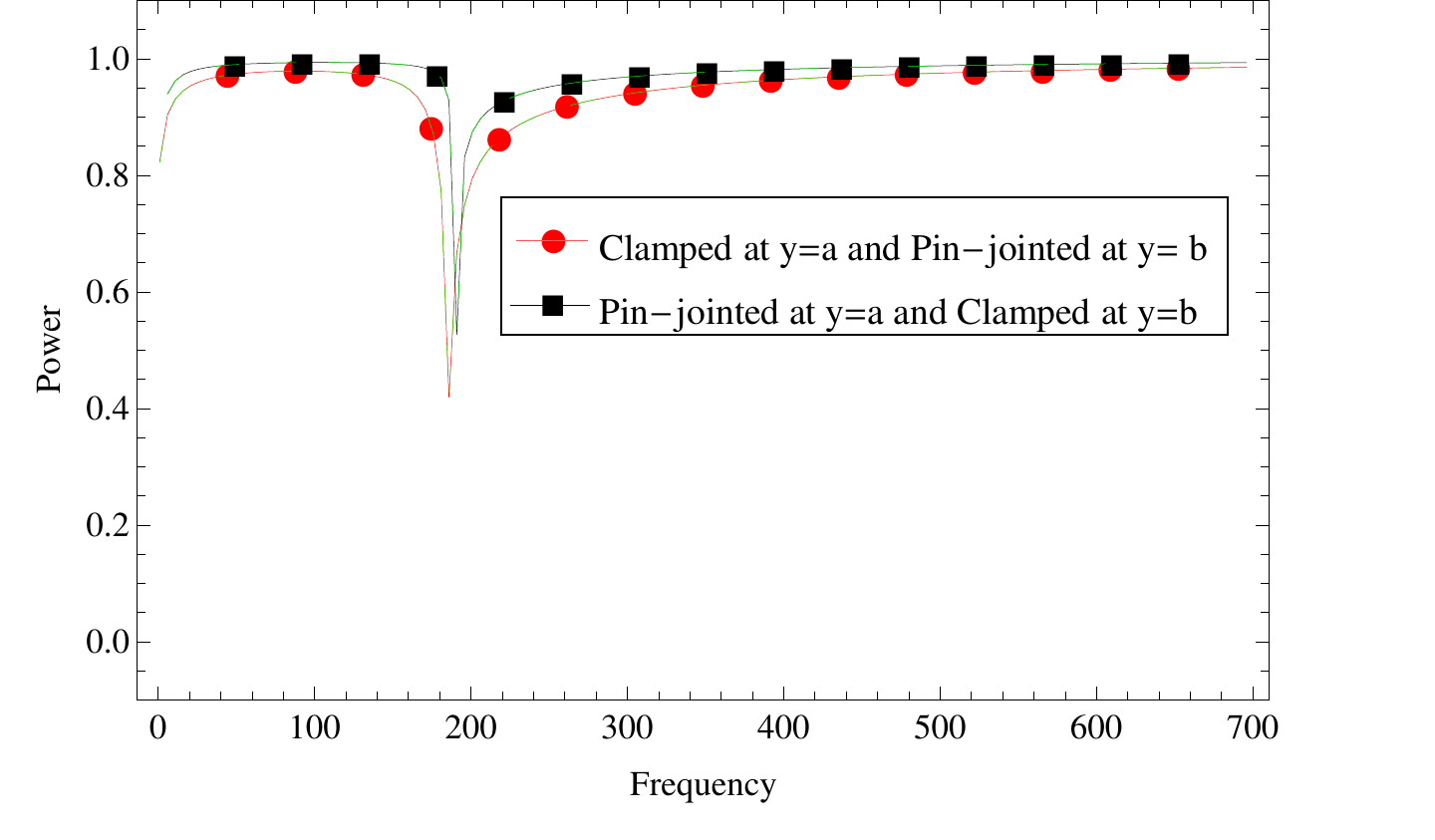}\label{fig:3c}}  
  	\subfigure[Mixed edge conditions with $N=25$, $\ell=1$.]{\includegraphics[width=.48\textwidth]{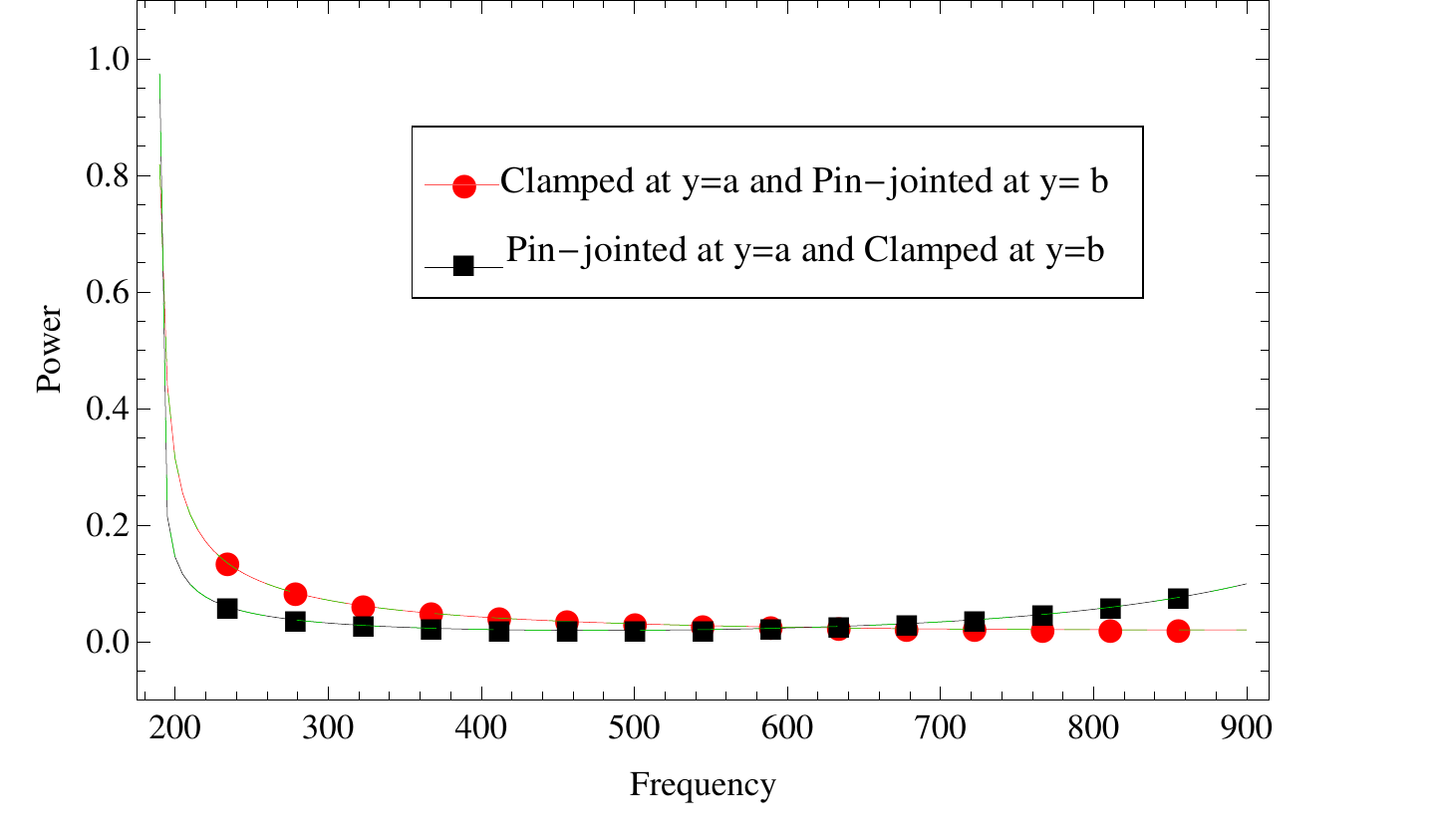}\label{fig:3d}}
  	\caption{Reflected power vs. frequency  using the MTG (with solid lines and markers $\blacksquare$, $\blacktriangle$, and $\CIRCLE$)  and  the MMA  (with dashed lines, without markers). Left: structure-borne fundamental  mode  incidence.  Right: fluid-borne second mode incidence.}
  	\label{fig:3}
\end{figure}

\begin{figure}[htb!]
\subfigure[Same edge conditions with $N=40$, $\ell=0$]{\includegraphics[width=.48\textwidth]{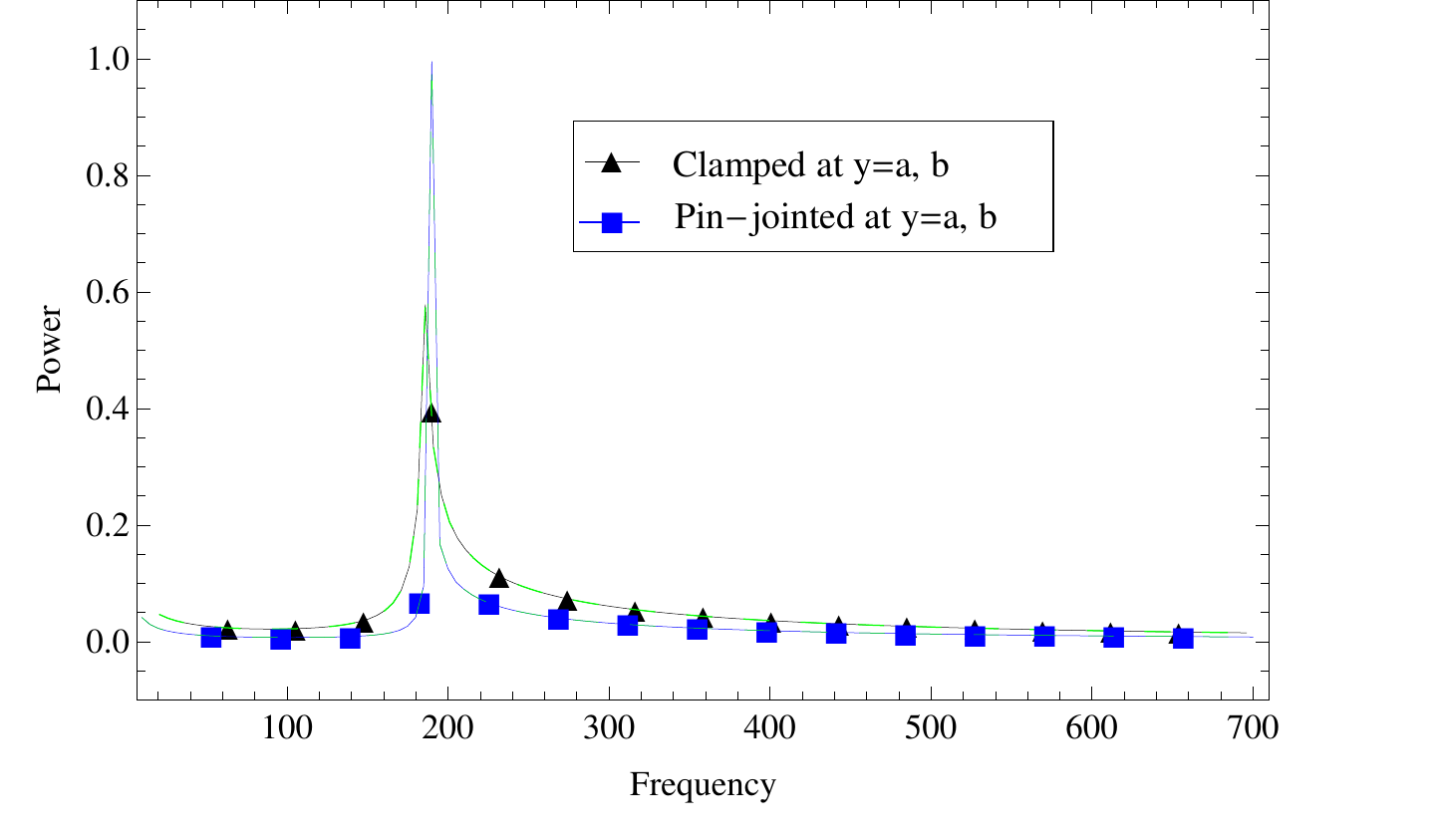}\label{fig:4a}}  
\subfigure[Same edge conditions with $N=40$, $\ell=1$]{\includegraphics[width=.48\textwidth]{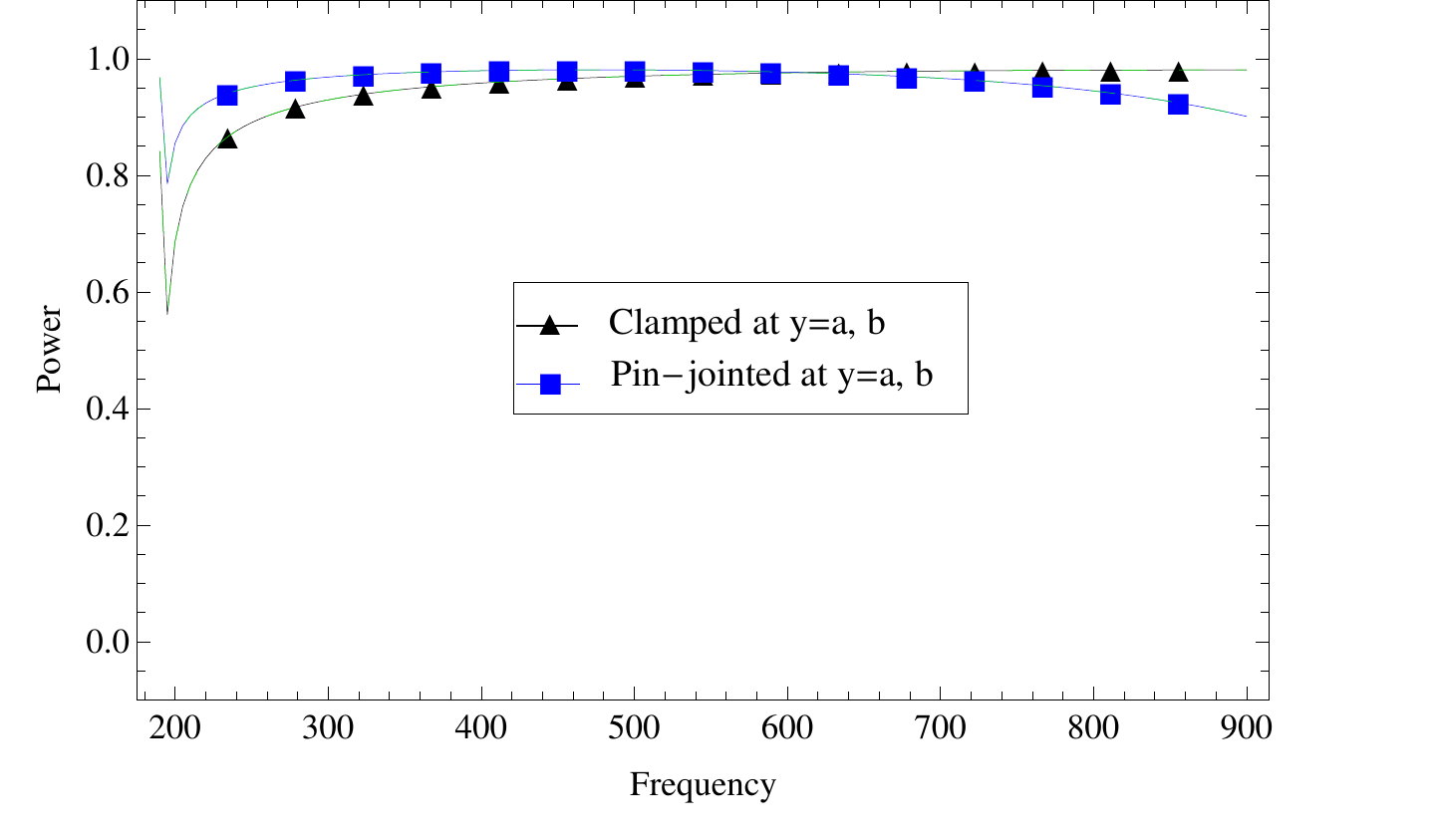}\label{fig:4b}}
\\
\subfigure[Mixed edge conditions with $N=25$, $\ell=0$]{\includegraphics[width=.48\textwidth]{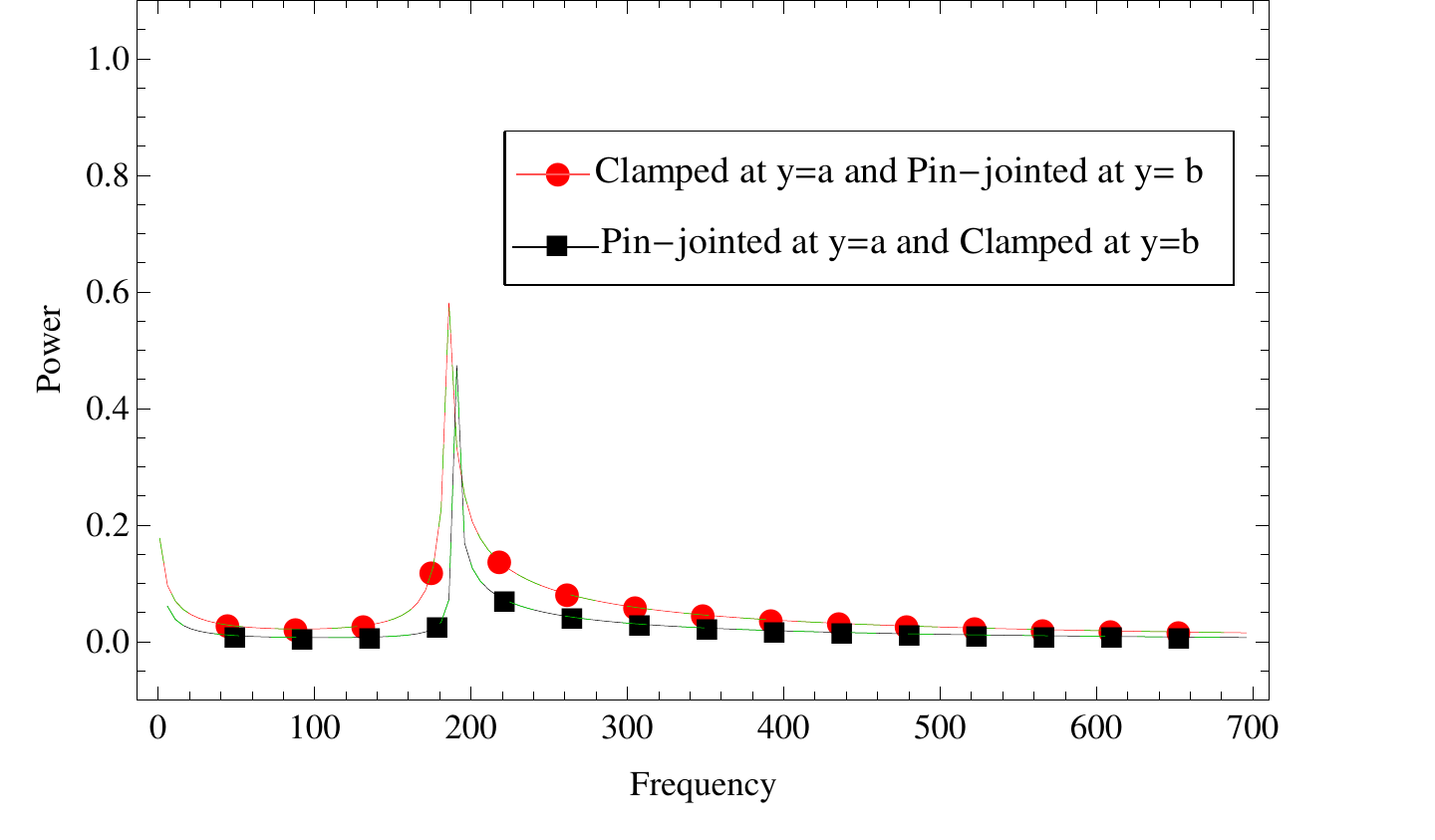}\label{fig:4c}}  
\subfigure[Mixed edge conditions with $N=25$, $\ell=1$]{\includegraphics[width=.48\textwidth]{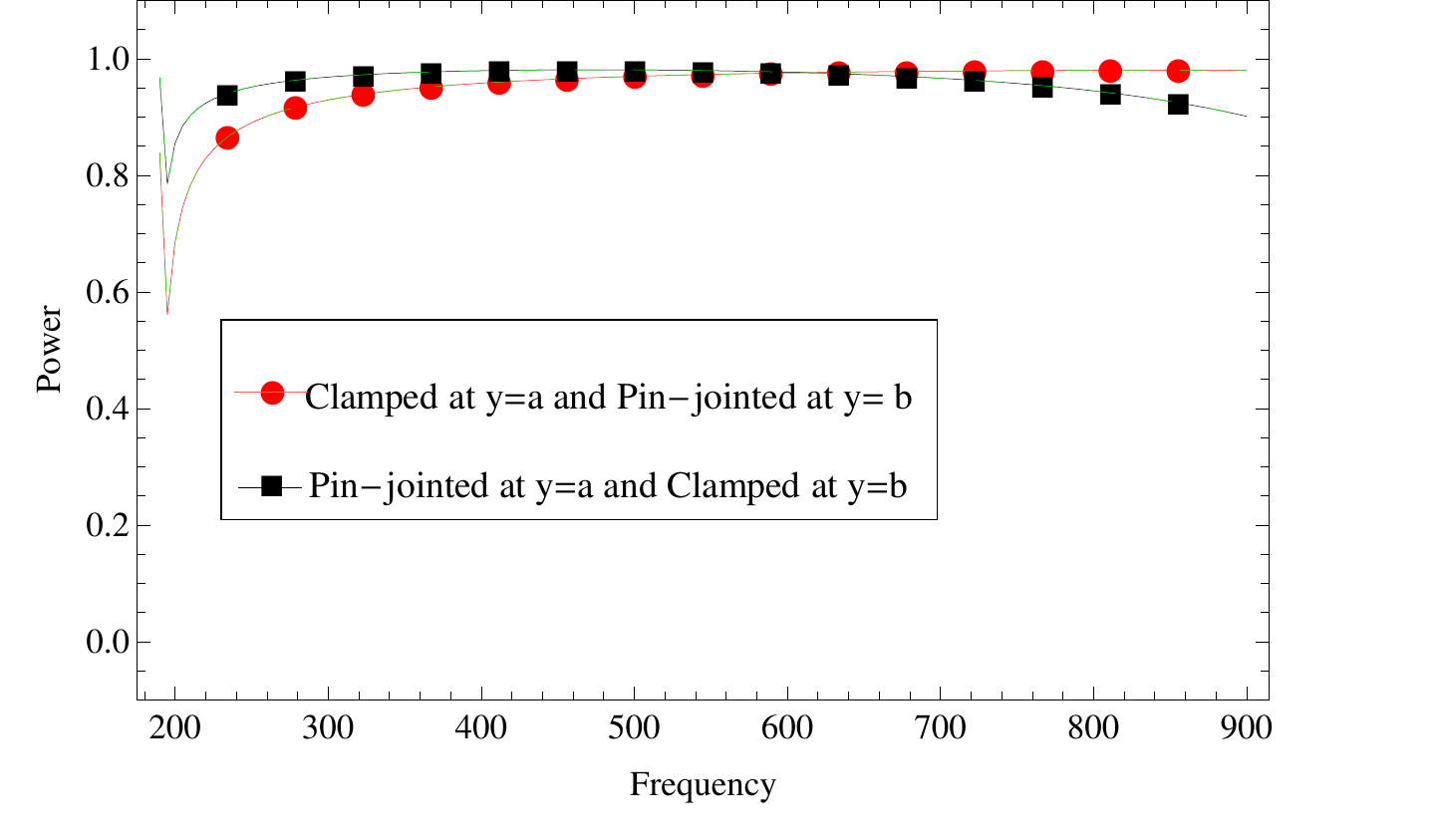}\label{fig:4d}}
\caption{Transmitted power vs. frequency using the MTG technique  (with solid lines and markers $\blacksquare$, $\blacktriangle$, and $\CIRCLE$)  and the MMA (with dashed lines, without markers). Left:  structure-borne fundamental mode of incidence. Right: fluid-borne second mode of incidence.}
	\label{fig:4}
\end{figure}

Figures \ref{fig:3}-\ref{fig:4} furnish, respectively, the plots for the reflected and transmitted powers versus frequency obtained by truncating and inverting the MTG and the MMA linear systems obtained in Sections \ref{sec:TG} and \ref{sec:Modal}. A good agreement between the scattering energy curves furnished by the MTG and the MMA in all frequency bands can be observed. 
  
The results in Figs. \ref{fig:3a}-\ref{fig:3b} correspond to the horizontal and vertical elastic plates with both clamped or pin-joint connections. Clearly, the series solutions for the structure-borne fundamental mode incidence ($ell=0$) converge faster than the fluid-borne second mode incidence ($ell=1$). The effects of clamped and pin-joint connections become apparent when the second cut-on mode of the inlet/outlet starts propagating. Observe that the maximum radiated energy goes on reflection at $f=1$Hz. However, the amount of reflection decreases with frequency and reaches its minimum at the point of initiation of the second mode of the chamber cavity. Remark that the fundamental mode of extended inlet/outlet is always cut-on for the entire frequency band due to the presence of the zero eigenvalue. That renders a plane acoustic wave, a mode that always exists in a duct with a rigid boundary. The other propagating modes appear at higher frequencies. For example, the second cut-on mode of the extended inlet/outlet arises at $f=2881$Hz, however, it lies outside the frequency band considered herein. Similarly, the waveguide containing a cavity, whose lower wall is rigid but the upper walls consist of elastic plates, always has a fundamental duct mode throughout the bandwidth due to the presence of a real eigenvalue. The cut-on frequencies of the second and third modes of the duct containing a flexible cavity are $f=1161$Hz and $f=2301$Hz, respectively, which affect the scattering energies. A list of the cut-on frequencies is provided in Table \ref{table:1}. The results in Figs. \ref{fig:3c}-\ref{fig:3d} correspond to the effects of mixed edge connections of the horizontal and vertical plates (i.e., one is clamped and the other is pin-jointed) on the power reflection at different frequencies. Observe that the influence of the mixed edge conditions on the energy reflection is prominent, especially, when the second mode of inlet/outlet starts propagating at $f=191$Hz. Moreover, most of the energy is reflected for a fundamental mode incidence (see Fig. \ref{fig:3c}), in contrast to a fluid-borne mode instance for which energy is not much reflected (see Fig. \ref{fig:3d}). Apparently, the series solution converges more rapidly for the structure-borne mode incidence (see Fig.~\ref{fig:3c}) than the fluid-borne mode incidence (see Fig. \ref{fig:3d}).
\begin{table}[htb!]
	\centering
	\begin{tabular}{|c|c|c|}
		\hline
		Cut-on frequency $f$ in Hz	     & Extended inlet/outlet  \qquad    &   Chamber cavity \qquad      \\ \hline
		191  & 1                    & 1                        \\ \hline
		1161 & 1                    & 2                        \\ \hline
		2301 & 1                    & 3                        \\ \hline
		2881 & 2                    & 3                        \\ \hline
		3446 & 2                    & 4                        \\ \hline
		3996 & 2                    & 5                        \\ \hline
		4015 & 3                    & 5                        \\ \hline
	\end{tabular}
	\caption{Cut-on modes}\label{table:1}
\end{table}  

Figure \ref{fig:4} delineates the transmitted powers versus frequency with the same edge connections (Figs. \ref{fig:4a}-\ref{fig:4b}) and different edge connections (Figs. \ref{fig:4c}-\ref{fig:4d}) of the horizontal and vertical elastic plates. As with energy reflection, the edge conditions significantly influence energy transmission, especially when the first cut-on mode of inlet/outlet or the higher-order cavity modes start propagating. The maximum energy goes on transmission for fluid-borne mode incidence, and a negligible amount of scattering energy goes on transmission for structure-borne mode incidence.

\subsubsection{Transmission loss}
 
\begin{figure}[htb!]
	\subfigure[Same edge conditions with $N=40$, $\ell=0$]{\includegraphics[width=.48\textwidth]{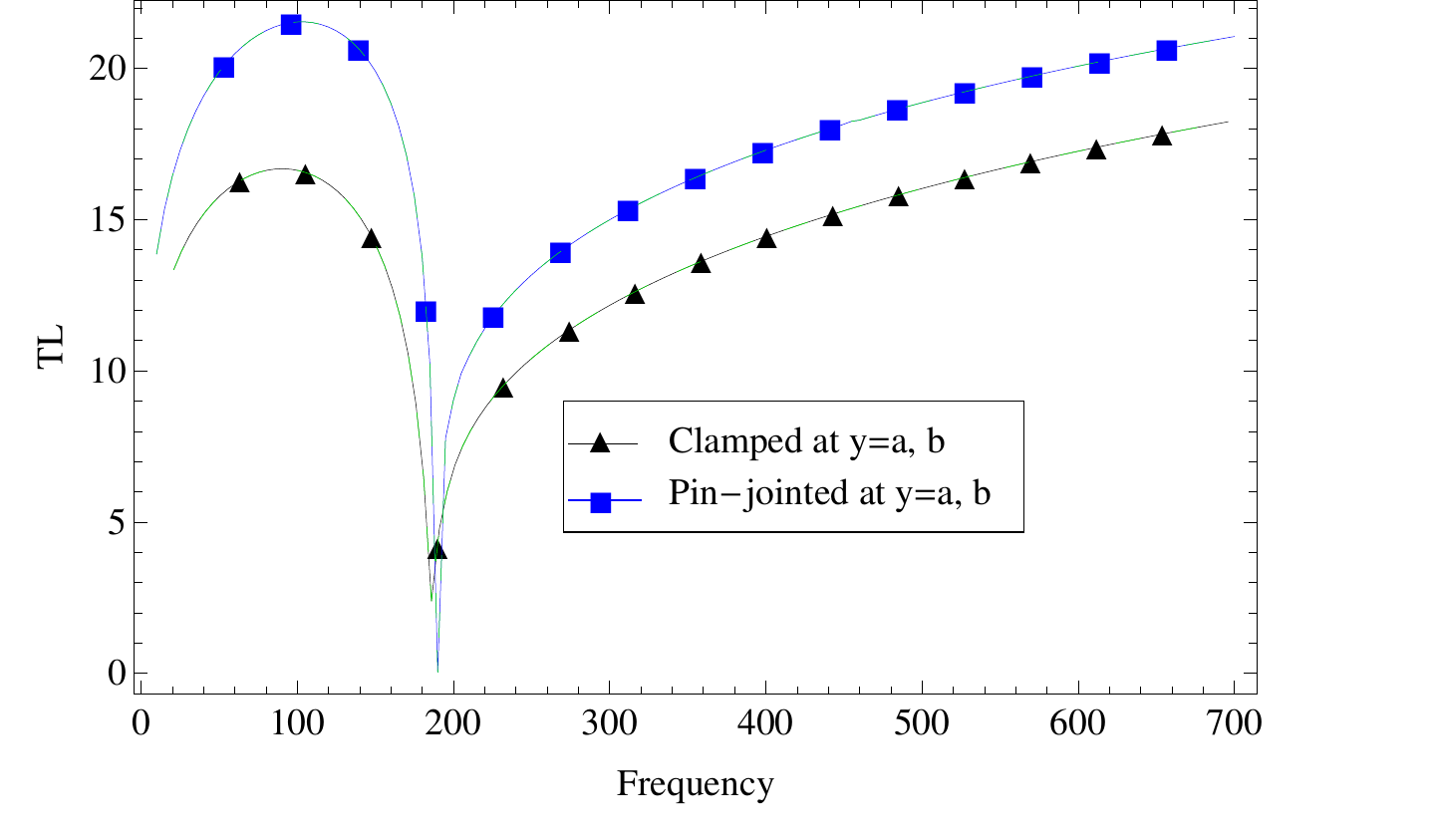}\label{fig:5a}}  
	\subfigure[Same edge conditions with $N=20$, $\ell=1$]{\includegraphics[width=.48\textwidth]{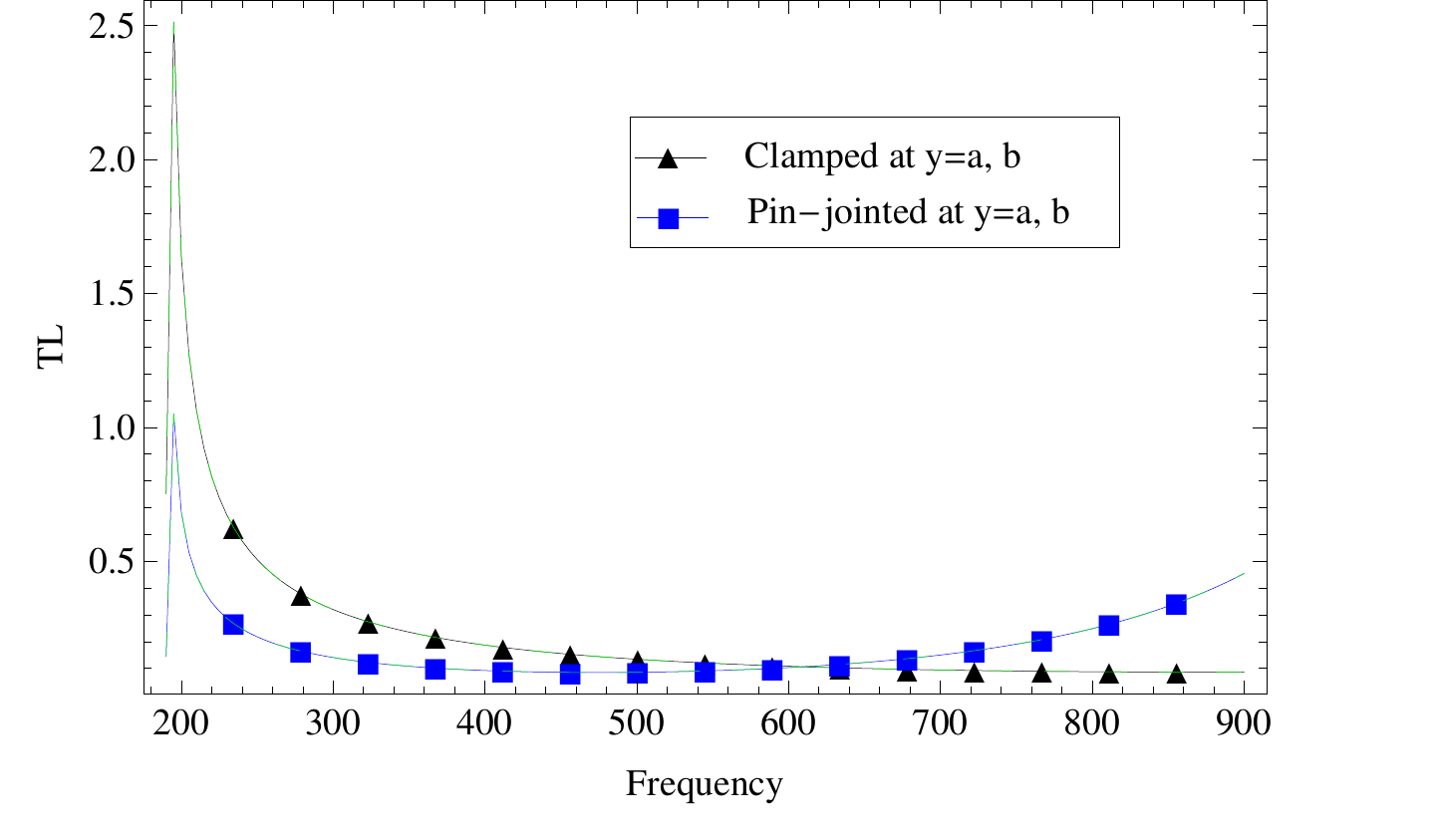}\label{fig:5b}}
	\\
\subfigure[Mixed edge conditions with $N=20$, $\ell=0$]{\includegraphics[width=.48\textwidth]{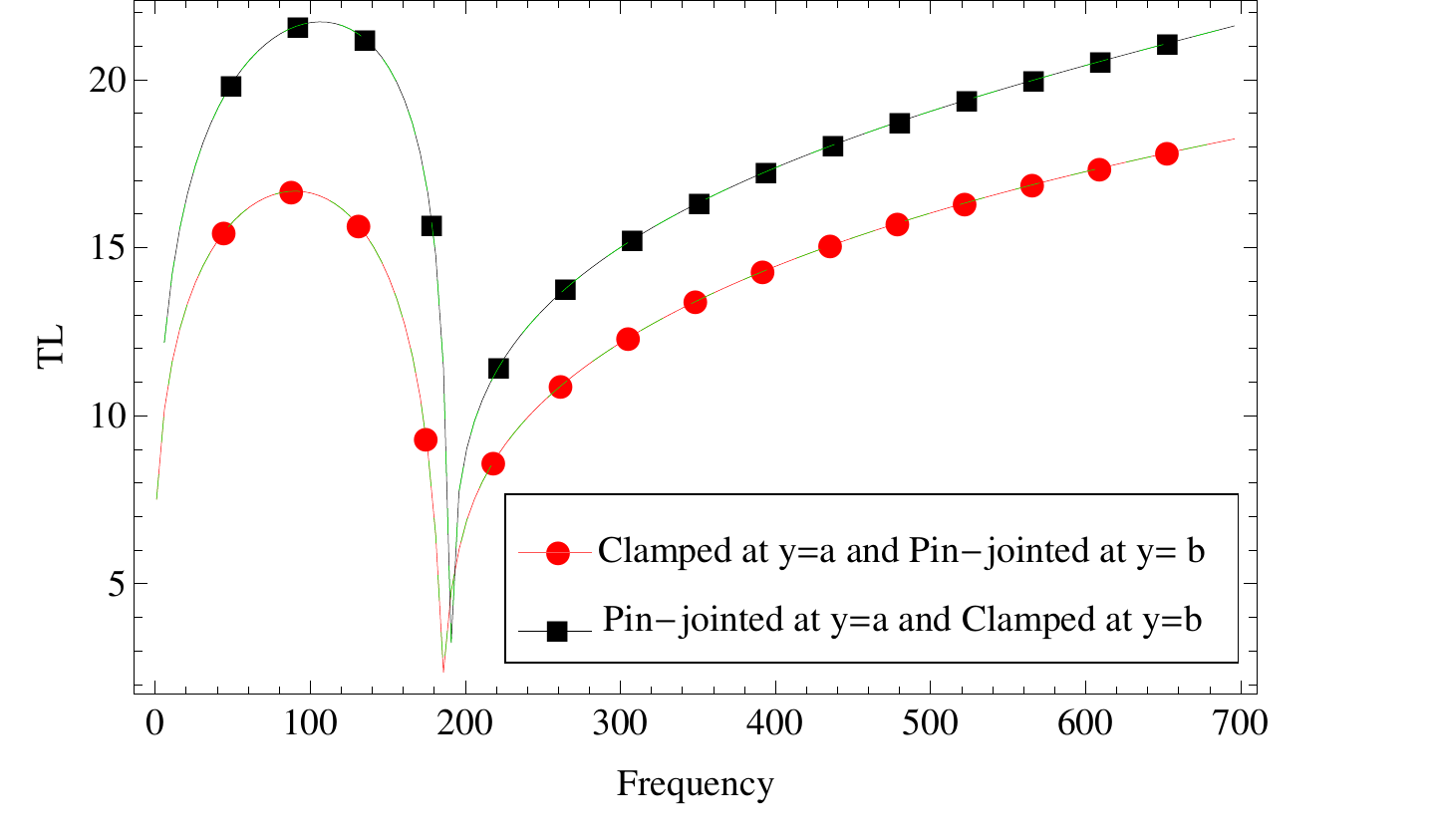}\label{fig:5c}}  
\subfigure[Mixed edge conditions with $N=20$, $\ell=1$]{\includegraphics[width=.48\textwidth]{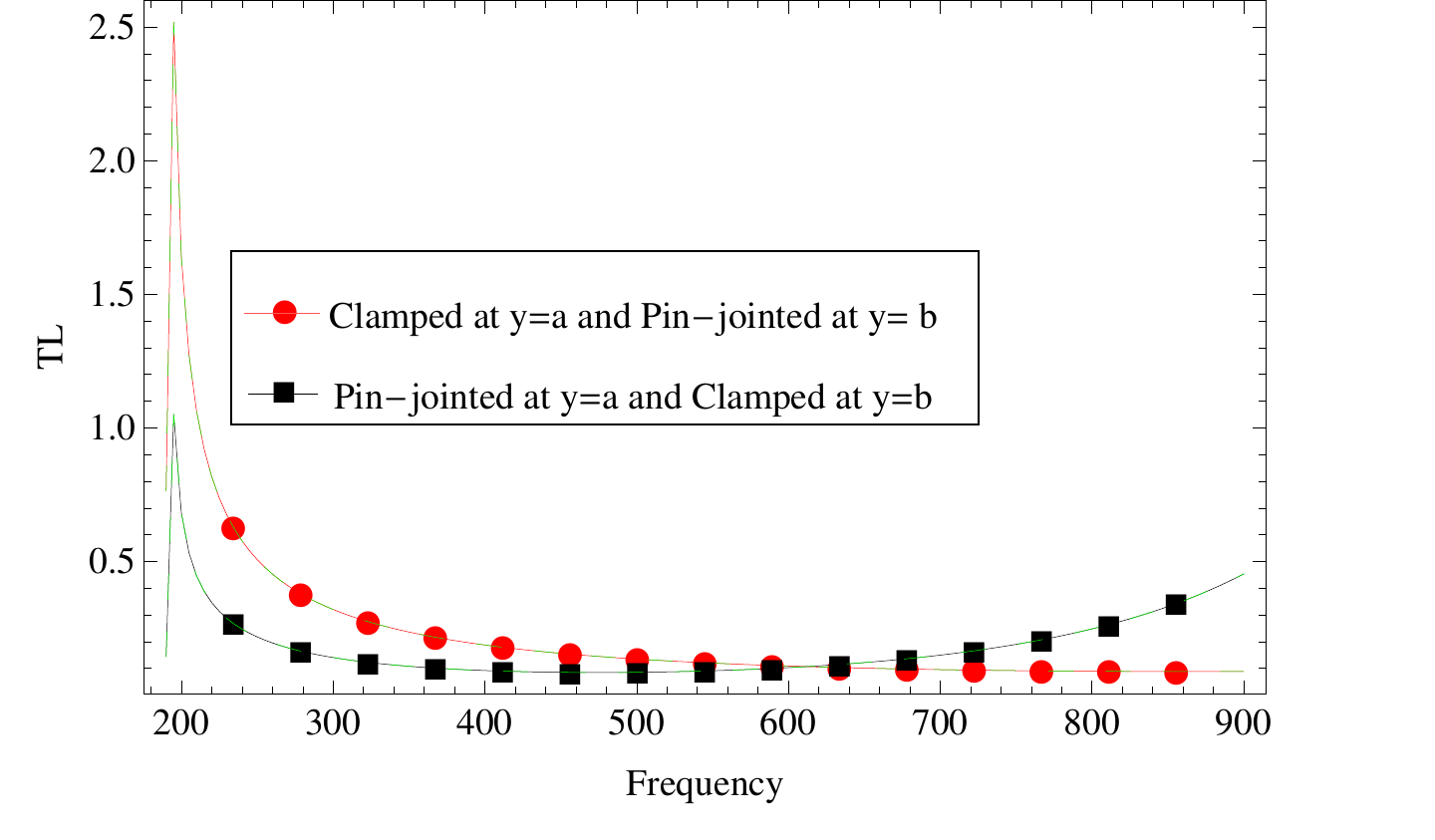}\label{fig:5d}}
\caption{Transmission  loss  vs. frequency using the MTG technique  (with solid lines and markers $\blacksquare$, $\blacktriangle$, and $\CIRCLE$)  and  the MMA (with dashed lines, without markers). Left:  structure-borne fundamental mode of incidence. Right: fluid-borne  second mode of incidence.}
\label{fig:5}
\end{figure}

Figure \ref{fig:5} elaborates on the transmission loss with respect to frequency in the presence of the expansion chamber for different sets of edge conditions on the horizontal and vertical elastic plates. Figures \ref{fig:5a}-\ref{fig:5b} correspond with all clamped or all pin-joint elastic plates. It is observed that maximum energy propagates along the structure and a stop-band is produced in the regime $2{\rm Hz}\leq f \leq 190{\rm Hz}$ as the next mode of inlet duct becomes cut-on at $f=191$Hz. This is a fluid-borne mode. When both modes start propagating, the transmission loss starts increasing with frequency. In other words, we get more reflection and absorption with the participation of additional modes. Interestingly, more leakage in the compressional waves is observed for the pin-joint connections than for the clamped edges due to zero bending moments at the edges. Therefore, transmission loss can be optimized by choosing pin-joint connections for the plates.

The transmission loss with respect to the frequency corresponding to different sets of edge connections of the horizontal and vertical elastic plates is provided in Figs. \ref{fig:5c}-\ref{fig:5d}. Once again, maximum energy propagates along the structure and a stop-band is produced in the frequency band $2{\rm Hz}\leq f \leq 190{\rm Hz}$. For different edge connections, more leakage in the compressional waves and, consequently, increased transmission loss are observed for the pin-joint-clamped structure (pin-joint at $y=a$, clamped at $y=b$) than for the clamped-pin-joint structure (clamped at $y=a$, pin-joint at $y=b$). 

\section{Conclusions}\label{sec:Con} 

The mode-matching-tailored-Galerkin approach and the mode-matching-modal approach have been discussed in this article to model fluid-structure couple wave propagation in a flexible waveguide containing bridging elastic plates with different edge conditions. 
%The formulation of these methods is based on the requirement of modeling a bridging elastic plate response subject to different edge conditions.
The displacement of a bridging elastic plate is assumed to satisfy the plate equation in the tailored-Galerkin approach, so that its homogeneous part determines the plate modes and its integral part links the flexible chamber vibration. The homogeneous part involves 8 auxiliary constants that are determined from the conditions on the edges of elastic plates. This article considers clamped and/or pin-jointed edges, but its scope is broad; we can consider restraint or any other type of connection depending on the physical conditions on elastic plates at edges. In the modal approach, the displacement of the bridging elastic plate is projected onto the eigenspace of the cavity region. In flexible ducts, where the fluid-structure coupled modes exist, the solution comprises non-orthogonal eigenfunctions, and the application of generalized orthogonal properties is essential. The coupling quantities are integrated to determine the auxiliary constants needed for the unique solution. The presence of auxiliary constants is linked to the connections on the finite edges of elastic plates and helps us incorporate physical constraints. On the other hand, these constants help us derive a set of linearly independent eigenfunctions from a set of linearly dependent eigenfunctions and ensure the uniqueness of the solution as well.  A minor drawback of the modal approach is that it requires more computational time to converge than the tailored-Galerkin approach. The tailred-Glaerkin approach outperforms previous approaches in that it does not require the implementation of additional algorithms to incorporate more practical edge conditions such as spring-like or restrained edge conditions. The formulation of plate displacements achieved via the tailored-Galerkin approach or the modal approach is incorporated in the matching conditions of normal velocities at the interfaces. The proposed schemes are validated by verifying the matching conditions at the interfaces and the conservation of power. The numerical results show that the variation of conditions at the joints has a significant impact on scattering energies and transmission loss.

\appendix
\section{Derivations}
\subsection{Derivations of  Eqs. \eqref{eq:19} and \eqref{eq:23}} \label{append:a}
Substituting  expressions \eqref{eq:11aa}, \eqref{eq:12}, and \eqref{eq:13aa} of $\phi_1$, $\phi_2$, and $\phi_3$  into the pressure flux conditions \eqref{eq:21a} and  \eqref{eq:21aa}, and subsequently using generalized orthogonality relation \eqref{eq:8},  we get 
\begin{align}
A_m=& - F_\ell\delta_{m,\ell}+{\Delta_{1,m} }\left(e_1+(\tau^2_m+2)e_2\right)+\dfrac{\alpha}{ \Gamma_{1,m}}\sum_{n=0}^{\infty}   R_{m,n}
\left(B_{n}e^{-is_{n}L}  +  C_{n}e^{is_{n}L}\right),
\label{eq:19s1}
\\
D_{m}= &\Delta_{1,m} \left(e_3+(\tau^2_m+2) e_4\right) +\dfrac{\alpha}{  \Gamma_{1,m}}\sum_{n=0}^\infty   R_{m,n}
\left(B_ne^{is_n L}  +  C_ne^{-is_{n}L}\right),
\label{eq:20s1}
\end{align}
where $\Delta_{\jmath,m}$, $R_{m,n}$, and $e_1,\cdots,e_4$  are defined in Eqs. \eqref{eq:9}, \eqref{eq:r1}, and \eqref{e14}, respectively.  Adding and subtracting Eqs. \eqref{eq:19s1}  and  \eqref{eq:20s1}, we arrive at
\begin{align}
A_m+D_m=& - F_{\ell}\delta_{m,\ell}+{\Delta_{1,m}} \left[ (e_1+e_3)+(\tau^2_m+2)  (e_2+e_4)\right]  
\nonumber  
\\
&-\dfrac{2 i \alpha}{\Gamma_{1,m}}\sum_{n=0}^{\infty}    R_{m,n}  \Pi_n^-  \left(B_n+  C_n\right).
\label{eq:19s11}
\\
A_m-D_m=& - F_{\ell}\delta_{m,\ell}+{\Delta_{1,m}} \left[(e_1-e_3)+(\tau^2_m+2)  (e_2-e_4)\right]  
\nonumber \\
&-\dfrac{2 i \alpha}{\Gamma_{1,m}}\sum_{n=0}^{\infty}   R_{m,n}  \Pi_n^+  \left(B_{n}-C_n\right).
\label{eq:19s111}
\end{align}
Finally, using the definitions \eqref{chi-psi} and \eqref{eq:r1} of $\Psi^{\pm}_m$,  $\chi^{\pm}_n$, $R_{m,n}$, and {$U_j^{\pm}$,  we get Eq. \eqref{eq:19}.

Similarly,  substituting Eqs. \eqref{eq:11aa}, \eqref{eq:12}, and \eqref{eq:13aa}  into Eqs. \eqref{eq:14aa}  and  \eqref{eq:14aaa},  and then using orthogonality relation \eqref{eq:8},  we obtain
\begin{align}
B_me^{-is_mL}  &-  C_me^{is_mL}=\dfrac{\Delta_{2,m}}{s_m}\left[e_5+(\gamma_m^2+2)  e_6\right] +\dfrac{\alpha}{s_m \Gamma_{2,m}} \left(F_\ell \xi_\ell R_{\ell,m}-\sum_{n=0}^\infty  R_{n,m}\xi_n A_n\right) 
\nonumber  \\
&-\dfrac{i\alpha}{s_m\Gamma_{2,m}}\sum_{j=1}^4 a_{j} A_{j,m}
 +\dfrac{i \alpha^2}{s_m \Gamma_{2,m}} \sum_{n=0}^{\infty} \dfrac{ T_{m,n}(B_{n}e^{-is_{n}L}+C_{n}e^{is_{n}L})}{\gamma_n^4-\mu^4},
\label{eq:23s2}    
\\
B_m e^{is_mL}&-C_me^{-is_mL}=\dfrac{\Delta_{2,m}}{s_m}\left[e_7+(\gamma_m^2+2)  e_8\right] +\dfrac{\alpha}{s_m \Gamma_{2,m}} \sum_{n=0}^\infty  R_{n,m}\xi_n D_n 
\nonumber \\
&-\dfrac{i\alpha }{s_m \Gamma_{2,m}}\sum_{j=1}^4a_{j+4} A_{j,m}-\dfrac{i \alpha^2}{s_m \Gamma_{2,m}} \sum_{n=0}^\infty\dfrac{T_{m,n}(B_ne^{is_{n}L}+C_ne^{-is_nL})}{\gamma_n^4-\mu^4}.\label{eq:24s2}    
\end{align}
Subtracting and adding Eqs.  \eqref{eq:23s2}  and  \eqref{eq:24s2}, we get 
\begin{align}
B_m+C_m=&\dfrac{i\Delta_{2,m}}{2s_m\Pi_m^+}\left[(e_5-e_7)+(\gamma^2_m+2)  (e_6-e_8)\right]
\nonumber \\
&+\dfrac{i \alpha}{2 s_m \Gamma_{2,m}\Pi_m^+}\left(F_\ell \xi_\ell  R_{\ell,m}-\sum_{n=0}^{\infty} \xi_n R_{n,m}(A_n+D_n)\right)
\nonumber  
\\  
&+\dfrac{\alpha}{2s_m  \Gamma_{2,m}\Pi_m^+}\sum_{j=1}^4(a_j-a_{j+4}) A_{j,m}
\nonumber
\\
&+\dfrac{i\alpha^2 }{s_m\Gamma_{2,m}\Pi_m^+}\sum_{n=0}^{\infty} \dfrac{\Pi_n^-T_{m,n}(B_{n}+C_{n})}{\gamma_n^4-\mu^4},
\label{eq:23s22}    
\end{align}
and 
\begin{align}
B_m-C_m=&\dfrac{i\Delta_{2,m}}{2s_m\Pi_m^-}\left[(e_5+e_7)+(\gamma^2_m+2)  (e_6+e_8)\right]
\nonumber 
 \\
&+\dfrac{i \alpha}{2 s_m \Gamma_{2,m}\Pi_m^-}\left(F_\ell \xi_\ell R_{\ell,m}-\sum_{n=0}^{\infty}\xi_n  R_{n,m}(A_n-D_n)\right)
\nonumber  
\\  
&+\dfrac{\alpha }{2s_m  \Gamma_{2,m}\Pi_m^-}\sum_{j=1}^4(a_j+a_{j+4}) A_{j,m}
\nonumber
\\
&+\dfrac{i\alpha^2}{s_m  \Gamma_{2,m}\Pi_m^-}\sum_{n=0}^\infty\dfrac{\Pi_n^+ T_{m,n}(B_n-C_n)}{\gamma_n^4-\mu^4}.
\label{eq:23s21}    
\end{align}
Finally, using the definitions \eqref{chi-psi}, \eqref{eq:r1}, and \eqref{eq:47a}, of $\chi^{\pm}_n$, $R_{m,n}$, $T_{m,n}$, and $V_j^{\pm}$, we get the expression \eqref{eq:23}.

\subsection{Derivations of Eqs. \eqref{eq:29q} and \eqref{eq:291q}}\label{appendix:b}

To determine  $U_{1}^{\pm}$,  we multiply Eq. \eqref{eq:19} with  $\gamma_m\sinh(\gamma_ma)$ and use the definition of $\Delta_{1,m}$  to get 
\begin{align}
\xi_{m} \Delta_{1,m}\Gamma_{1,m}\Psi^\pm_m=& - \xi_{m} \Delta_{1,m} F_\ell\delta_{m,\ell} \Gamma_{1,\ell}+\xi_{m} \Delta^2_{1,m}\Gamma_{1,m} U_{1}^\pm-2i\alpha\xi_{m} \Delta_{1,m}\sum_{n=0}^\infty \Pi^\mp_n R_{m,n}  \chi^\pm_n.
\label{eq:19nc}
\end{align}
On the other hand,  using  expressions \eqref{eq:11aa} and \eqref{eq:13aa} for $\phi_1$ and $\phi_3$ into edge conditions \eqref{eq:38a}, we see that
\begin{align*}
F_{\ell}\xi_{\ell} {\Delta_{1,\ell}\Gamma_{1,\ell}}-\sum_{m=0}^{\infty}{\xi_{m} \Delta_{1,m}\Gamma_{1,m}}A_m=0
\quad\text{and}\quad
\sum_{m=0}^{\infty}{\xi_{m} \Delta_{1,m}\Gamma_{1,m}}D_m=0.
\end{align*}
Consequently, 
\begin{align}
\sum_{m=0}^{\infty}{\xi_{m} \Delta_{1,m}\Gamma_{1,m}}\Psi^{\pm}_m=F_{\ell}{\xi_{\ell} \Delta_{1,\ell}\Gamma_{1,\ell}}.
\label{appeq:3}
\end{align}
Finally, expression \eqref{eq:28} is found by summing over $m$ in Eq. \eqref{eq:19nc} and then, using the relation \eqref{appeq:3}.

To determine $V_{1}^{\pm}$ we proceed in a similar fashion. Towards this end, we multiply Eq. \eqref{eq:23} by $\Delta_{2,m}\Gamma_{2,m}\Pi_m^{\pm}$, sum over $m$, impose edge conditions \eqref{eq:39} to finally arrive at Eqs. \eqref{eq:29q} and \eqref{eq:291q}.

\subsection{Derivation  of  Eq. \eqref{eq:23M}}\label{appendix:c}

To obtain the equations of continuity of velocity flux for the modal approach in the form of scattering coefficients, we substitute Eqs. \eqref{eq:11aa}, \eqref{eq:12}, \eqref{eq:13aa}, \eqref{5311a}, and \eqref{5312} into Eqs. \eqref{eq:14aa} and \eqref{eq:14aaa}. After normalizing resultant with the aid of orthogonality relation  \eqref{eq:8} and simplifying, we arrive at
\begin{align}
B_me^{-is_mL}-C_me^{is_m L}=&{\dfrac{\Delta_{2,m}}{s_m}}\left[e_5+(\gamma^2_m+2)  e_6\right]+\dfrac{\alpha}{s_m \Gamma_{2,m}} \left(F_\ell \xi_\ell R_{\ell,m}-\sum_{n=0}^{\infty}  R_{n,m}\xi_n A_n\right) 
\nonumber\\
&+\dfrac{i\alpha }{s_m  \Gamma_{2,m}}\left(\sum_{j=1}^8L_j \Phi^j_m+\alpha^2\sum_{n,q=0}^{\infty}\dfrac{(B_ne^{-is_nL}+C_ne^{is_nL})T_{m,q}T_{n,q}}{\Gamma_{2,q}(\gamma_q^4-\mu^4)}\right),
\label{eq:23s3}    
\\
B_me^{is_mL}- C_me^{-is_m L}=&\dfrac{\Delta_{2,m}}{s_m}\left[e_7+(\gamma^2_m+2)  e_8\right] +\dfrac{\alpha}{s_m \Gamma_{2,m}}\sum_{n=0}^{\infty} R_{n,m}\xi_n D_n
\nonumber \\
&+\dfrac{i\alpha }{s_m  \Gamma_{2,m}}\left(\sum_{j=1}^8M_j \Phi^j_m-\alpha^2\sum_{n,q=0}^{\infty}\dfrac{(B_ne^{+is_nL}+C_ne^{-s_nL})T_{m,q}T_{n,q}}{\Gamma_{2,q}(\gamma_q^4-\mu^4)}\right).\label{eq:24s3}    
\end{align}
Subtracting and adding of Eqs. \eqref{eq:23s3} and \eqref{eq:24s3} then lead to
\begin{align}
B_m+C_m=&\dfrac{i\Delta_{2,m}}{2s_m\Pi_m^+}\left[(e_5-e_7)+(\gamma^2_m+2)  (e_6-e_8)\right]
\nonumber 
\\
&+\dfrac{i \alpha}{2 s_m \Gamma_{2,m}\Pi_m^+}\left(F_\ell \xi_\ell  R_{\ell,m}-\sum_{n=0}^{\infty}  R_{n,m}\xi_{\ell}(A_n+D_n)\right) 
\nonumber  
\\  
&-\dfrac{\alpha }{2s_m  \Gamma_{2,m}\Pi_m^+}\sum_{j=1}^8(L_j-M_j) \Phi^j_m
\nonumber
\\
&+\dfrac{2i\alpha^3 }{2s_m  \Gamma_{2,m}\Pi_m^+} \sum_{n,q=0}^{\infty}\dfrac{ T_{m,q}T_{n,q}\Pi_n^-(B_n+C_n)}{\Gamma_{2,q}(\gamma_q^4-\mu^4)},
\label{eq:23s4}    
\end{align}
and 
\begin{align}
B_m-C_m=&\dfrac{i\Delta_{2,m}}{2s_m\Pi_m^-}\left[(e_5+e_7)+(\gamma^2_m+2)  (e_6+e_8)\right]
\nonumber 
\\
&+\dfrac{ i\alpha}{2 s_m \Gamma_{2,m}\Pi_m^-}\left(F_\ell \xi_\ell  R_{m,\ell}-\sum_{n=0}^\infty  R_{m,n}\xi_\ell(A_n-D_n)\right)
\nonumber  
\\  
&-\dfrac{\alpha }{2s_m  \Gamma_{2,m}\Pi_m^-}\sum_{j=1}^8(L_j+M_j) \Phi^j_m
\nonumber
\\
&+\dfrac{2i\alpha^3 }{2s_m  \Gamma_{2,m}\Pi_m^-} \sum_{n,q=0}^{\infty}\dfrac{ T_{m,q}T_{n,q}\Pi_n^+(B_n-C_n)}{\Gamma_{2,q}(\gamma_q^4-\mu^4)}.
\label{eq:23s5}    
\end{align}
Finally, using the definitions \eqref{chi-psi}, \eqref{eq:r1}, and \eqref{eq:47a}, of $\chi^{\pm}_n$, $R_{m,n}$, $T_{m,n}$, $V_1^{\pm}$, and  $V_2^{\pm}$, we arrive at the expression \eqref{eq:23M}.

\section*{Competing interests}

All the authors declare that they have no competing interests.

\section*{Acknowledgment}

A.W. would like to acknowledge the financial supported by the Nazarbayev University, Kazakhstan through Faculty Development Competitive Research Grant Program (FDCRGP) under Grant 1022021FD2914. N.A. would like to acknowledge financial support from the Gulf University for Science and Technology for an Internal Seed Grant (No. 278877).


\begin{thebibliography}{10}
\bibitem{Doak}
P.~E. Doak.
\newblock Excitation, transmission and radiation of sound from source
  distributions in hard-walled ducts of finite length (i): The effects of duct
  cross-section geometry and source distribution space-time pattern.
\newblock {\em Journal of Sound and Vibration}, 31(1):1--72, 1973.

\bibitem{Dowell}
E.~H. Dowell, G.~F. Gorman, and D.~A. Smith.
\newblock Acoustoelasticity: General theory, acoustic natural modes and forced
  response to sinusoidal excitation, including comparisons with experiment.
\newblock {\em Journal of Sound and Vibration}, 52(4):519--542, 1977.

\bibitem{Pan}
J.~Pan and D.~A. Bies.
\newblock The effect of fluid–structural coupling on sound waves in an
  enclosure—theoretical part.
\newblock {\em The Journal of the Acoustical Society of America},
  87(2):691--707, 1990.

\bibitem{Huang}
L.~Huang.
\newblock A theoretical study of duct noise control by flexible panels.
\newblock {\em The Journal of the Acoustical Society of America},
  106(4):1801--1809, 1999.

\bibitem{ammari1}
H.~Ammari, G.~Bao, and A.~W. Wood.
\newblock An integral equation method for the electromagnetic scattering from
  cavities.
\newblock {\em Mathematical Methods in the Applied Sciences},
  23(12):1057--1072, 2000.

\bibitem{ammari2}
H.~Ammari, G.~Bao, and A.~W. Wood.
\newblock Analysis of the electromagnetic scattering from a cavity.
\newblock {\em Japan Journal of Industrial and Applied Mathematics},
  19:301--310, 2002.

\bibitem{Ming}
R.~Ming and J.~Pan.
\newblock Insertion loss of an acoustic enclosure.
\newblock {\em The Journal of the Acoustical Society of America},
  116(6):3453--3459, 2004.

\bibitem{Lau1}
R.~Kirby.
\newblock Modeling sound propagation in acoustic waveguides using a hybrid
  numerical method.
\newblock {\em The Journal of the Acoustical Society of America},
  124(4):1930--1940, 2008.

\bibitem{Du}
J.~T. Du, W.~L. Li, H.~A. Xu, and Z.~G. Liu.
\newblock Vibro-acoustic analysis of a rectangular cavity bounded by a flexible
  panel with elastically restrained edges.
\newblock {\em The Journal of the Acoustical Society of America},
  131(4):2799--2810, 2012.

\bibitem{Lawrie12}
J.~B. Lawrie.
\newblock On acoustic propagation in three-dimensional rectangular ducts with
  flexible walls and porous linings.
\newblock {\em The Journal of the Acoustical Society of America},
  131(3):1890--1901, 2012.

\bibitem{Kim}
H.-S. Kim, S.-R. Kim, S.-H. Lee, Y.-H. Seo, and P.-S. Ma.
\newblock Sound transmission loss of double plates with an air cavity between
  them in a rigid duct.
\newblock {\em The Journal of the Acoustical Society of America},
  139(5):2324--2333, 2016.

\bibitem{Mimani}
A.~Mimani.
\newblock {\em Acoustic Analysis and Design of Short Elliptical End-Chamber
  Mufflers}.
\newblock Springer, Singapore, 2021.

\bibitem{Norris}
A.~N. Norris and G.~R. Wickham.
\newblock Acoustic diffraction from the junction of two flat plates.
\newblock {\em Proceedings of the Royal Society of London. Series A:
  Mathematical and Physical Sciences}, 451(1943):631--655, 1995.

\bibitem{Cannell}
P.~A. Cannell and J.~T. Stuart.
\newblock Edge scattering of aerodynamic sound by a lightly loaded elastic
  half-plane.
\newblock {\em Proceedings of the Royal Society of London. A. Mathematical and
  Physical Sciences}, 347(1649):213--238, 1975.

\bibitem{Lawrie}
J.~B. Lawrie and I.~D. Abrahams.
\newblock Scattering of fluid loaded elastic plate waves at the vertex of a
  wedge of arbitrary angle, i: analytic solution.
\newblock {\em IMA Journal of Applied Mathematics}, 59(1):1--23, 1997.

\bibitem{Grant}
A.~D. Grant and J.~B. Lawrie.
\newblock Propagation of fluid-loaded structural waves along a duct with
  smoothly varying bending characteristics.
\newblock {\em Quarterly Journal of Mechanics and Applied Mathematics},
  53(2):299--321, 2000.

\bibitem{Thompson}
I.~Thompson, I.~D. Abrahams, and A.~N. Norris.
\newblock On the existence of flexural edge waves on thin orthotropic plates.
\newblock {\em The Journal of the Acoustical Society of America},
  112(5):1756--1765, 2002.

\bibitem{Warren}
D.~P. Warren, J.~B. Lawrie, and I.~M. Mohamed.
\newblock Acoustic scattering in waveguides that are discontinuous in geometry
  and material property.
\newblock {\em Wave Motion}, 36(2):119--142, 2002.

\bibitem{Law1999}
J.~B. Lawrie and I.~D. Abrahams.
\newblock An orthogonality relation for a class of problems with high-order
  boundary conditions; applications in sound structure interaction.
\newblock {\em Quarterly Journal of Mechanics and Applied Mathematics},
  52(2):161--181, 1999.

\bibitem{Lawrie2007}
J.~B. Lawrie.
\newblock On eigenfunction expansions associated with wave propagation along
  ducts with wave-bearing boundaries.
\newblock {\em IMA Journal of Applied Mathematics}, 72(3):376--394, 2007.

\bibitem{Hassan09}
M.~Ul-Hassan, M.~H. Meylan, and M.~A. Peter.
\newblock Water-wave scattering by submerged elastic plates.
\newblock {\em Quarterly Journal of Mechanics and Applied Mathematics},
  62(3):321--344, 2009.

\bibitem{Afzal16}
M.~Afzal, R.~Nawaz, and A.~Ullah.
\newblock Attenuation of dissipative device involving coupled wave scattering
  and change in material properties.
\newblock {\em Applied Mathematics and Computation}, 290:154--163, 2016.

\bibitem{Afzal16b}
M.~Afzal, M.~Ayub, R.~Nawaz, and A.~Wahab.
\newblock Mode-matching solution of a scattering problem in flexible waveguide
  with abrupt geometric changes.
\newblock In {\em Imaging, Multi-scale and High Contrast Partial Differential
  Equations}, volume 660, page 113. American Mathematical Society, Providence,
  USA, 2016.

\bibitem{Nawaz21}
T.~Nawaz, M.~Afzal, and A.~Wahab.
\newblock Scattering analysis of a flexible trifurcated lined waveguide
  structure with step-discontinuities.
\newblock 96(11):115004, 2021.

\bibitem{Afzal21}
M.~Afzal and J.~U. Satti.
\newblock The traveling wave formulation of a splitting chamber containing
  reactive components.
\newblock 91:1959--1980, 2021.

\bibitem{Bilal22}
H.~Bilal and M.~Afzal.
\newblock On the extension of the mode-matching procedure for modeling a
  wave-bearing cavity.
\newblock {\em Mathematics and Mechanics of Solids}, 27(2):348 -- 367, 2021.

\bibitem{Aqsa}
A.~Yaseen and R.~Nawaz.
\newblock Acoustic radiation through a flexible shell in a bifurcated circular
  waveguide.
\newblock {\em Mathematical Methods in the Applied Sciences}, 10.1002/mma.8902.

\bibitem{Sumbul}
M.~Ul-Hassan, M.~H. Meylan, A.~Bashir, and Sumbul M.
\newblock Mode matching analysis for wave scattering in triple and
  pentafurcated spaced ducts.
\newblock {\em Mathematical Methods in the Applied Sciences},
  39(11):3043--3057, 2016.

\bibitem{RN13Jasa}
R.~Nawaz and J.~B. Lawrie.
\newblock Scattering of a fluid-structure coupled wave at a flanged junction
  between two flexible waveguides.
\newblock {\em The Journal of the Acoustical Society of America},
  134(3):1939--1949, 2013.

\bibitem{Afzal20Jasa}
M.~Afzal and S.~Shafique.
\newblock Attenuation analysis of flexural modes with absorbent lined flanges
  and different edge conditions.
\newblock {\em The Journal of the Acoustical Society of America},
  148(1):85--99, 2020.

\bibitem{AfzalCNSNS21}
M.~Afzal, S.~Shafique, and A.~Wahab.
\newblock Analysis of traveling waveform of flexible waveguides containing
  absorbent material along flanged junctions.
\newblock {\em Communications in Nonlinear Science and Numerical Simulation},
  97:105737, 2021.

\bibitem{Afzal22Jasa}
M.~Afzal, J.~U. Satti, A.~Wahab, and R.~Nawaz.
\newblock Scattering analysis of a partitioned membrane-bounded cavity with
  material contrast.
\newblock {\em The Journal of the Acoustical Society of America},
  151(1):31--44, 2022.

\bibitem{LawrieAfzal}
J.~B. Lawrie and M.~Afzal.
\newblock Acoustic scattering in a waveguide with a height discontinuity
  bridged by a membrane: a tailored {G}alerkin approach.
\newblock {\em Journal of Engineering Mathematics}, 105:99--115, 2017.

\bibitem{lawrie2012comments}
J.~B. Lawrie.
\newblock Comments on a class of orthogonality relations relevant to
  fluid-structure interaction.
\newblock {\em Meccanica}, 47:783--788, 2012.

\bibitem{lawrie2006tuning}
J.~B. Lawrie and I.~M.~M. Guled.
\newblock On tuning a reactive silencer by varying the position of an internal
  membrane.
\newblock {\em The Journal of the Acoustical Society of America},
  120(2):780--790, 2006.

\end{thebibliography}
\end{document}